\documentclass[prx,notitlepage,longbibliography]{revtex4-2}

\pdfoutput=1

\usepackage{amsmath,amssymb,bm} 
\usepackage{graphicx}
\usepackage[papersize={8.5in,11in}]{geometry}
\usepackage[tight]{subfigure} 

\usepackage{color}
\definecolor{darkblue}{rgb}{0.,0.,0.4}
\definecolor{darkred}{rgb}{0.5,0.,0.}
\definecolor{BlueViolet}{RGB}{138,43,226}
\definecolor{SkyBlue}{RGB}{30,144,255}
\definecolor{DarkGreen}{RGB}{0,100,0}
\usepackage[pdftex,colorlinks=true,linkcolor=darkblue,citecolor=blue,urlcolor=darkred]{hyperref}
\usepackage{makecell}

\usepackage{amsfonts,bbold}
\usepackage{upgreek}
\usepackage{slashed}
\usepackage{latexsym}
\usepackage{ulem}

\usepackage{wasysym}			
\usepackage{amsthm}
\usepackage{xcolor}

\newcommand{\nn}{\nonumber \\}
\newcommand{\be}{\begin{equation}}
\newcommand{\ee}{\end{equation}}

\renewcommand{\vec}[1]{{\bf #1}}

\renewcommand{\epsilon}{\varepsilon}

\begin{document}
\title{Robust Marginal Fermi Liquid In Birefringent Semimetals}

\author{Ipsita Mandal}
\affiliation{Institute of Nuclear Physics, Polish Academy of Sciences, 31-342 Krak\'{o}w, Poland}
\affiliation{Department of Physics, Stockholm University, AlbaNova University Center,
106 91 Stockholm, Sweden}

\begin{abstract}
We investigate the interplay of Coulomb interactions and correlated disorder in pseudospin-3/2 semimetals, which exhibit birefringent spectra in the absence of interactions. Coulomb interactions drive the system to a marginal Fermi liquid, both for the two-dimensional (2d) and three-dimensional (3d) cases. Short-ranged correlated disorder in 2d, or a power-law correlated disorder 3d, has the same engineering dimension as the Coulomb term, in a renormalization group (RG) sense.
In order to analyze the combined effects of these two kinds of interactions, we apply a dimensional regularization scheme and derive the RG flow equations. The results show that the marginal Fermi liquid phase is robust against disorder.
\end{abstract}

\maketitle
\tableofcontents

\section{Introduction}

Quasiparticles with pseudospin-3/2 and having a birefringent linear spectrum with two distinct Fermi velocities, can be realized from simple tight-binding models 
in both two-dimensional (2d) and three-dimensional (3d) systems.
Examples in 2d include decorated $\pi$-flux square lattice \cite{malcolm,prb.85.235119,prb.90.045131}, honeycomb lattices \cite{prb84.195422,watanabe_2011}, and shaken optical lattices \cite{prb.84.165115,prl.107.253001}. The 3d counterparts are captured in various systems having strong spin-orbit coupling
\cite{bradlyn,prb.94.195205}, such as the antiperovskite family \cite{PhysRevB.90.081112} (with the chemical formula A$_3$BX) and the CaAgBi-family materials with a stuffed Wurtzite structure \cite{PhysRevMaterials.1.044201}.
We consider the low-energy effective Hamiltonian of such semimetals and investigate the phases resulting from the interplay of Coulomb interactions and correlated disorder.

In both 2d and 3d, Coulomb interactions drive a clean system (i.e. without disorder) into a marginal Fermi liquid phase \cite{malcolm-bitan}, which we rederive here in order to correct some minor algebraic factors in the loop calculations of Ref.~\cite{malcolm-bitan}, and also to set up our minimal subtraction scheme to obtain the renormalized action. Next, we add disorder to the system. In 2d, a short-ranged correlated disorder has the same engineering dimension as the Coulomb term, in a renormalization group (RG) sense. However, in 3d, this has a lower scaling dimension (and hence it is an irrelevant operator), and that is why we consider a power-law correlated disorder with the same scaling dimension as the Coulomb terms.
Since the added disorder and Coulomb interactions have the same scaling dimension, we treat them on an equal footing in the RG scheme.
By implementing a controlled $\epsilon$-expansion below the upper critical dimension, we derive the RG flow equations at leading order (i.e. by employing the corrections coming from the logarithmically divergent one-loop Feynman diagrams), which are valid when Coulomb interactions and disorder are both weak. The stable fixed points of the coupled differential equations show that the marginal Fermi liquid phase is not destroyed by disorder. This is in contrast with other four-band semimetal phases considered earlier \cite{rahul-sid,ips-rahul}. As a result, the birefringent spectrum pseudospin-3/2 semimetals are a promising platform to observe the putative marginal Fermi liquid phases arising at band-touching points.

The paper is organized as follows. In Sec.~\ref{secmodel}, we describe the system in 2d and 3d, in the presence of Coulomb interactions. In Sec.~\ref{seccoulomb}, we show the emergence of the marginal Fermi liquid phase due to the Coulomb terms.
In Sec.~\ref{secdis}, we add disorder and compute the one-loop diagrams.
In Sec.~\ref{secrg}, we derive the coupled RG equations due to the interplay of Coulomb interactions and disorder. The fixed points of the RG equations and their stability indicate the resulting phases. Finally, we end with a summary and outlook in Sec.~\ref{conclude}. 

\section{Model}
\label{secmodel}

\begin{figure}[]
\begin{center}
\includegraphics[width=0.2 \textwidth]{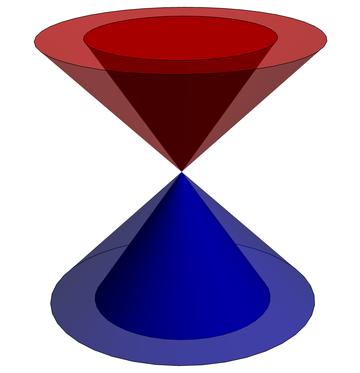}
\caption{\label{figdisp}
Birefringent semimetals show conical energy dispersions with two different slopes, when the energy eigenvalues are plotted against the $k_x$-$k_y$ plane.}
\end{center}	
\end{figure}

The Hamiltonian describing the pseudospin-3/2 fermions in $d_0$ spatial dimensions is given by \cite{malcolm,malcolm-bitan}
\begin{align}
\label{eqham}
\mathcal{H}_0 (\mathbf k) = H_1(\mathbf k) +H_2(\mathbf k)\,, 
\quad H_1(\mathbf k) = v\sum_{j=1}^{d_0}\Gamma_{j0}\,k_j\,, 
\quad H_2(\mathbf k)= v\sum_{j=1}^{d_0}\zeta\,\Gamma_{0 j}\,k_j \,,\quad 
\Gamma_{\mu\nu} =\sigma_\mu \otimes \sigma_\nu\,,
\end{align}
where $\sigma_0$ is the $2\times 2$ identity matrix, $d_0$ takes the value $2$ (for 2d) or $3$ (for 3d),
and $\sigma_j$ are the Pauli matrices (with $j=1,2,3$). The four eigenvalues are given by $ \pm v \left( 1 \pm \zeta \right) |\mathbf k|$.
Hence, the emergent quasiparticles have birefringent spectra with linear dispersions (see Fig.~\ref{figdisp}). Here $\zeta$  is the birefringence parameter, with $|\zeta| < 1$.
In the following, we rescale the spatial momenta such that $v$ is equal to one.

The Coulomb interaction can be written as an effective four-fermion term, such that the total action takes the form: 
\begin{align}
& \mathcal{S}_0 =  \int \frac{dk_0 \,d^{d_0} {\mathbf k} } {(2\,\pi)^{d_0+1}}\, 
{\tilde \psi}^{\dag}(k_0,\mathbf k)
\left [ - i \,k_0 + \mathcal{H}_0(\mathbf k) \right]
 {\tilde \psi}  (k_0, \mathbf k) 
\nn & 
\qquad - e^2
\int \frac{dq_0\, dk_0 \,dk_0'\,
d^{d_0} {\mathbf q}\, d^{d_0} {\mathbf k}\, d^{d_0} {\mathbf k'}}
{(2\pi)^{3 \left (d_0+1 \right )}}\, 
V(|\mathbf q|)\,\tilde{\psi}^{\dag}  (k_0,\mathbf k)\,
\tilde{\psi}^{\dag} ( k_0' ,{\mathbf k}')\, 
\tilde{\psi}  (k_0+ q_0,{\mathbf k}+\mathbf q) \,
\tilde{\psi} (k_0'- q_0,{\mathbf k}'-\mathbf q)  \,,
\nn & V(|\mathbf q|) = \begin{cases}
\frac{1} { |\mathbf q|  } \text{ for } d_0=2 \\
\frac{1} { |\mathbf q|^{2} } \text{ for } d_0=3
\end{cases}.
\label{action}
\end{align}
in the momentum space. The tilde over $\psi$ indicates that it is the Fourier-transformed version. To avoid confusion between the actual (physical) spatial dimension, and the artificial spatial dimension used to perform a dimensional regularization (for a system of a given physical dimensionality), we have used use $d_0$ for the former and will use $d$ for the latter.
From the non-interacting part of the action, the bare Green's function is given by
\begin{align}
G_0 (k_0,\mathbf k)
= \frac{ \left[ i\,k_0 + H_1 (\mathbf k) + H_2 (\mathbf k) \right ]
\left[ k_0^2 +   \left( 1+\zeta^2 \right)\mathbf k^2 
- 2\, H_1 (\mathbf k)\, H_2 (\mathbf k) \right ]
}
{\left[ k_0^2 +   \left( 1+\zeta \right)^2\mathbf k^2 \right ]
\left[ k_0^2 +  \left( 1 - \zeta \right)^2\mathbf k^2 \right ]
}\,.
\end{align}

Let us determine the engineering dimensions of the fields and coupling constants at the non-interacting Gaussian fixed point ($ e^2 = 0$), by setting $[\mathbf{k}]= 1$ in the kinetic term.
Then, from the fermion dispersion, we get $[k_0]= 1 $. This leads to $[\tilde \psi] = -\frac{d}{2}- 1 $
and $[\zeta] = 0$. Finally, $[e^2] = \begin{cases}
2-d  & \text{ for } d_0=2 \\
3-d & \text{ for } d_0=3
\end{cases} \,,$ which means that the Coulomb interaction is marginal at the
upper critical dimension $d_c =d_0$.
We will employ the dimensional regularization scheme which involves continuing to
$\left( d_c - \epsilon \right) $ dimensions, while assuming that the angular and spinorial (matrix) structure remains the same as in $d_0$. In other words, the radial momentum integrals are performed
with respect to a $(d_c - \epsilon)$-dimensional measure $\int 
\frac{d^{d_c - \epsilon} \mathbf k}{(2\,\pi)^{d_c - \epsilon}}$.

\section{Marginal Fermi liquid in the clean system}
\label{seccoulomb}
 In this section, we will compute the one-loop corrections arising due to the Coulomb interactions, and determine the RG flow equations.
We introduce a mass scale $\mu$ to make the coupling constants dimensionless \cite{denis,ips-uv-ir1,ips-uv-ir2,ips-nfl-u1}. Therefore, in Eq.~\eqref{action}, we replace $e^2$ by $ e^2\,\mu^\epsilon$.

\subsection{One-loop contributions}

The one-loop contribution to the fermion self-energy due to the long-range Coulomb interaction is 
\begin{align}
\Sigma_c(k_0, \mathbf k)&= -\frac{e^2 \,H_1(\mathbf k)}
{ c_{d_0} \,\epsilon} \left( \frac{\mu} {|\mathbf k|} \right)^{\epsilon}
+\mathcal{O}(\epsilon^0) \,,
\text{where }
c_{d_0} = \begin{cases}
{4 \,\pi} & \text{ for } d_0=2 \\
{3\,\pi^2} & \text{ for } d_0 =3 
\end{cases}\,.
\end{align}
Ref.~\cite{malcolm-bitan} missed a factor of $2$ which comes from the two possible exchange contractions \footnote{
There are two possible contractions (matching one $\tilde \psi^{\dag}$ with one $\tilde \psi$) of the interaction term in Eq.~\eqref{action} that gives the fermion self-energy. In the real space,
the `Hartree' contractions are of the form $\langle \psi^{\dag} (t,x) 
\,\psi (t,x) \rangle \, \psi^{\dag} (t',{x'})\, \psi (t',{x'})$, and correspond to tadpole diagrams. These simply shift the overall chemical potential, and can be ignored (since we assume that the renormalized chemical potential is at the band-crossing point). However, the `exchange' contractions contribute (cannot be ignored). These latter contributions are of the form $\langle \psi^{\dag} (t,x) \,\psi(t',{x'})
\rangle \, \psi^{\dag}(t',{x'}) \,\psi(t,x)$, and $\langle \psi^{\dag} (t',x') \,\psi(t,{x}) \rangle \, \psi^{\dag}(t,{x}) \,\psi(t',x')$. Due to two ways of contractions, we get a factor of 2.
}.

Since $V(|\mathbf q|)$ is an analytic function of momentum in $d_0=3$, we need to compute the corrections to the Coulomb interaction coming from fermion loops. On the other hand, the Coulomb interaction does not receive any correction in $d_0=2$, as it is non-analytic, implying a non-local term in the position space. This is because a fermion loop can only yield corrections that are analytic in momentum. The absence of corrections in 2d is also clearly demonstrated by explicitly computing the loop-integrals.
\begin{enumerate}
\item First let us consider the ZS diagram, where the contractions of the fermion lines can be made in four distinct ways. Including the proper combinatorial factor, we get the correction
\begin{align}
\Pi_{cc}^{ZS} (q_0,\mathbf q)
= -\frac{4\, e^4 \,\mu^{2 \,\epsilon}} {2\,\mathbf q^4}
\int \frac{dk_0 \,d^{d_c-\epsilon} {\mathbf k} } 
{(2\,\pi)^{d_c+1-\varepsilon}}
\text{Tr} \left[ G_0(k_0,\mathbf k+ \mathbf q)\, G_0(k_0,\mathbf k) \right ]
=\frac{ e^4 \,\mu^\epsilon } 
{ 3\, \pi^2\,{\mathbf q^2}\,\epsilon}
\left( \frac{\mu} {|\mathbf q|} \right)^{\epsilon}
\,\delta_{d_0,3}
+\mathcal{O}(\epsilon^0)\,.
\end{align}

\item The second is the vertex correction (VC) diagram, which gives
\begin{align}
\Gamma_{cc}^{VC} (q_0,\mathbf q)
= \frac{ e^4 \,\mu^{2 \,\epsilon}} { \mathbf q^2 }
\int \frac{dk_0 \,d^{d_c-\epsilon} {\mathbf k} } 
{(2\,\pi)^{d_c+1-\varepsilon}}
\frac{ G_0(k_0,\mathbf k+ \mathbf q)\, G_0(k_0,\mathbf k) } {\mathbf k^2}
= 0+\mathcal{O}(\epsilon^0)\,.
\end{align}

\item Finally, we can show that the ZS$^\prime$ and BCS diagrams do not contribute to the clean-system RG flows, as the concerned loop-integrals are convergent.

\end{enumerate}

\subsection{RG equations using the minimal subtraction scheme}

For the clean system with Coulomb interactions, the counterterm action is given by
\begin{align}
\label{counter1} 
S^c_{clean} & =  \int \frac{dk_0 \,d^{d} {\mathbf k} } {(2\,\pi)^{d+1}}\, 
{\tilde \psi}^{\dag}(k_0,\mathbf k)
\left( -i \,A_1\,k_0 + A_2\, \Gamma_{j0}\,k_j
+ A_3\, \zeta \,\Gamma_{0j}\,k_j \right)
 {\tilde \psi}  (k_0, \mathbf k) 
\nn & 
\qquad - e^2\,\mu^{\varepsilon}
\int \frac{dq_0\, dk_0 \,dk_0'\,
d^d{\mathbf q}\, d^d {\mathbf k}\, d^d {\mathbf k'}}
{(2\pi)^{3(d+1)}}\, A_4\,
V(|\mathbf q|)\,\tilde{\psi}^{\dag}  (k_0,\mathbf k)\,
\tilde{\psi}^{\dag} ( k_0' ,{\mathbf k}')\, 
\tilde{\psi}  (k_0+ q_0,{\mathbf k}+\mathbf q) \,
\tilde{\psi} (k_0'- q_0,{\mathbf k}'-\mathbf q)  \,,\nn
A_{n} & \equiv \,Z_n-1 = 
\sum_{\lambda=1}^\infty \frac{Z_{ n,\lambda}}
{\epsilon^\lambda}  \text{ with }  n=1,2,3,4 \,,
\end{align} 
and $d=d_c-\varepsilon$.

Adding the counterterms to the original $\mathcal S_0 $, and denoting the bare quantities by the index ``B'', we obtain the renormalized action as:
\begin{align}
 \label{ren-action} 
 {\mathcal S}^{ren}_{clean} & =   \int \frac{d{k_0}_B \,
d^{d} {\mathbf k}_B } {(2\,\pi)^{d+1}}\, 
{\tilde \psi}_B^{\dag}({k_0}_B, {\mathbf k}_B)
\left[ -i \,{k_0}_B +  \Gamma_{j0}\,{k_j}_B
+  \zeta_B \,\Gamma_{0j}\,{k_j}_B \right]
 {\tilde \psi}  ({k_0}_B, \mathbf k_B) 
\nn & \quad - e_B^2\,\mu^{\varepsilon}
\int \frac{d{q_0}_B\, d{k_0}_B \,d{k_0'}_B\,
d^d{\mathbf q}_B\, d^d {\mathbf k}_B\, d^d {\mathbf k'}_B}
{(2\pi)^{3(d+1)}}\, 
V(|\mathbf q_B|)\,\tilde{\psi}_B^{\dag} ({k_0}_B,\mathbf k_B)\,
\nn & \hspace{ 2 cm }\times
\tilde{\psi}_B^{\dag} ( {k_0'}_B ,{\mathbf k}'_B)\, 
\tilde{\psi}_B ({k_0}_B + {q_0}_B ,{\mathbf k}_B+ {\mathbf q}_B) \,
\tilde{\psi}_B ({k_0'}_B- {q_0}_B,{\mathbf k}'_B-\mathbf q_B) \,.
\end{align}  

The bare and renormalized quantities are related by the following convention:
\begin{align}
\label{scale1}
& {k_0}_B= Z_1 \, k_0\,,\quad \mathbf k_B = {Z_2} \,\mathbf k \,, \quad  
\tilde \psi_B = Z_\psi^{1/2} \tilde \psi \,,\quad
Z_\psi  = \frac{1} { Z_1 \, Z_2^{2-\epsilon} }\,,\quad
\zeta_B = \frac{Z_3 \,\zeta}{ Z_2 }\,,\quad
e^2_B =\frac{ Z_4\,e^2\, \mu ^{\epsilon }}{ Z_1  \,Z_2^{1-\epsilon} } \,.
\end{align}
Note that if we had kept the velocity $v$ in Eq.~\eqref{action}, we would obtain $v_B=v$, which means that it does not flow under RG, which also justifies our setting $v=1$ at the outset.
To one-loop order, we have $Z_n= 1+\frac{Z_{n,1}} {\epsilon}$, where
\begin{align}
 \label{rgmy} 
Z_{1,1} &= 0 \,,\quad
 Z_{2,1}  =  -\frac{e^2 }
{c_{d_0} } \,,\quad
Z_{3,1}  =  0  \,,\quad
Z_{4,1}  = \frac{ e^2  } { 3\,\pi^2 }\,\delta_{d_0,3} \,.
\end{align}

Let us define:
\begin{align}
z = 1 -\frac{\partial \ln \left(\frac{ Z_2}{Z_1}\right )} {\partial \ln \mu}  \,,\quad 
\eta  =\frac{1}{2} \frac{\partial \ln Z_\psi}{\partial \ln \mu}\,,
\end{align}
where $z$ is the dynamical critical exponent, and $\eta$ is the anomalous dimension of the fermions.
Since the bare quantities do not depend on $\mu$, their total derivative with respect to $\mu$ should vanish. Therefore, $\frac{d \ln \zeta_B}{d \ln \mu}=0$ and $\frac{d \ln e^2_B}{d \ln \mu}=0$ give:
\begin{align}
\label{eqz}
\frac{\partial \zeta }{\partial \ln \mu}\equiv \beta_\zeta =
 \left(   \frac{\partial \ln Z_2 }{\partial \ln \mu}
-\frac{\partial \ln Z_3 }{\partial \ln \mu} \right ) \zeta\,,
 \text{ and  }
\frac{\partial e^2 }{\partial \ln \mu}\equiv \beta_e =
- \left[ \epsilon -\frac{\partial \ln Z_1 }{\partial \ln \mu}
-  \left( 1-\epsilon  \right) \frac{\partial \ln Z_2 }{\partial \ln \mu}
+\frac{\partial \ln Z_4 }{\partial \ln \mu} \right ] e^2 \,,
\end{align}
respectively.

Using the expansions
$z = z^{(0)} ,$ and $ \beta_\zeta =
\beta_\zeta^{(0)} + \epsilon \, \beta_\zeta^{(1)},$ and $ \beta_e =
\beta_e^{(0)} + \epsilon \, \beta_e^{(1)},$
and comparing the powers of $\epsilon$ from the regular (non-divergent) terms of the equations
\begin{align}
& Z_1\, Z_2\left( z-1 \right) =  
Z_2 \,\frac{\partial  Z_1 } {\partial \ln \mu}
- Z_1\,\frac{\partial Z_2 }{\partial \ln \mu}\,,\quad
Z_2\, Z_3\, \beta_\zeta =
 \left(  Z_3\, \frac{\partial  Z_2 }{\partial \ln \mu}
-Z_2\,\frac{\partial  Z_3 }{\partial \ln \mu} \right ) \zeta\,,\nn
&  Z_1\,Z_2 \,Z_4\,\beta_e =
- \left[ Z_1\,Z_2 \,Z_4\,\epsilon -
Z_2\,Z_4 \frac{\partial  Z_1 }{\partial \ln \mu}
-  Z_1\,Z_4\left( 1-\epsilon  \right) \frac{\partial  Z_2 }{\partial \ln \mu}
+ Z_1\,Z_2 \,\frac{\partial  Z_4 }{\partial \ln \mu} \right ] e^2\,.
\end{align}
we get:
\begin{align}
z=1-\frac{e^2}{ c_{d_0}}\,,\quad
\frac{d \zeta} {dl} = - \frac{e^2\,\zeta } { c_{d_0} } \,,\quad
\frac{de^2}{dl} = \begin{cases}
 \left(\epsilon  - \frac{e^2 } { 4\,\pi } \right) e^2 
& \text{ for } d_0=2 \\
\left ( \epsilon - \frac{ 2\,e^2 } { 3\,\pi^2 } \right) e^2 
& \text{ for } d_0=3
\end{cases} \,,
\end{align}
where $l= -\ln \mu $ is the RG flow parameter (or the floating length scale).

There are two fixed points: (1) the Gaussian fixed point at which $\zeta =0 $, $e^2=0$, $z=1$, and $\eta=0$; and (2) the marginal Fermi liquid fixed point at which $\zeta =0 $,
$e^2 = \begin{cases}
4 \,\pi \,  \epsilon & \text{ for } d_0=2 \\
\frac{3\, \pi^2  \,\epsilon }{2} 
& \text{ for } d_0= 3
\end{cases} \, ,$
$ z = \begin{cases}
1- \epsilon & \text{ for } d_0=2 \\  
1 - \frac{\epsilon }{2} & \text{ for } d_0= 3
\end{cases} \, ,$ and
$\eta= \begin{cases}
- \epsilon & \text{ for } d_0=2 \\  
 - \frac{3\,\epsilon }{4} & \text{ for } d_0= 3
\end{cases} \,.$
The former is unstable, while the latter is a stable fixed point. This can be confirmed by writing down the linearized flow equations in the vicinity of the fixed points, and computing the stability matrix. This matrix has all negative eigenvalues for the marginal Fermi liquid fixed point.

\section{Addition of disorder}
\label{secdis}


We now consider the effect of correlated disorder on the non-interacting system.
We find that the engineering dimension for a short-ranged disorder coupling is given by $2-d_0+\epsilon$. Therefore, disorder is marginal in 2d, whereas irrelevant in 3d.
Therefore, in 3d, it is expected that the marginal Fermi liquid behaviour will be unchanged up to a critical strength of the disorder. Hence, we will focus on $d_0=2$ for short-ranged correlated disorder. On the other hand, we focus on power-law correlated disorder for $d_0=3$, such that the disorder realizations have a zero mean and a disorder correlation function proportional to $\frac{1}{|\mathbf k|}$ in the momentum space. This power-law correlated disorder has a coupling constant with the engineering dimension $3-d_0+\epsilon $, which is the same as the Coulomb interactions in $d_0=3$. We note that even when considering initial conditions for the RG with only long-range correlated disorder, short-range correlated disorder is generated perturbatively already at one-loop order, and should be kept in the space of couplings. By contrast, long-range correlated disorder cannot be generated perturbatively from short-range correlated disorder.

We now consider the 2d and 3d cases in two separate subsections. We will consider a minimum set of disorder realizations to keep the calculations simple.

\subsection{2d case}

If we consider disorder which is a scalar both in the spinor and position spaces (i.e. $\Gamma_{00}$), then we find that the loop corrections generate disorder terms having all possible matrix structures in the spinor space, captured by the generic matrix $\Gamma_{\mu \nu}$. Hence, we need to keep all these couplings from the start. This makes the calculation unwieldy. In order to keep it simple, while still being to extract the essential physics, we will then set $\zeta =0 $ and analyze the system for which the two velocities coincide. This seems to be a reasonable simplification to employ, as for the clean system with Coulomb interactions, $\zeta $ is marginal and indeed flows to zero under RG.

For the $\zeta =0$ Hamiltonian, the minimal set of disorder couplings we need to consider is given by
\begin{align}
\mathcal{S}^{2d}_{\text{dis}} & = -\mu^{\epsilon }  \sum \limits_{a,b=1}^n 
 \int d\tau\, d\tau' \,d^{2-\epsilon} \mathbf x \,\Big[\,
W_0\left( \psi_a^{\dag} \, {\psi_a}\right)_{\mathbf x,\tau}
 \left ( \psi_b^{\dag} \, \psi_b \right )_{\mathbf x,\tau'}
+ W_1 \left( \psi_a^{\dag} \,\Gamma_{30}\, {\psi_a}\right)_{\mathbf x,\tau}
\left ( \psi_b^{\dag} \,\Gamma_{30}\, \psi_b \right )_{\mathbf x,\tau'} \nn
& \hspace{ 4.5 cm}
+ W_2 \sum \limits_{j=1,2}
\left( \psi_a^{\dag} \,\Gamma_{j0}\, {\psi_a}\right)_{\mathbf x,\tau}
\left ( \psi_b^{\dag} \,\Gamma_{j0}\, \psi_b \right )_{\mathbf x,\tau'} \Big ]\, ,
\end{align}
after disorder-averaging using the replica trick. Here, the replica indices have been indicated by $a$ and $b$, which run over $n$ replicas of the fermion fields. The limit $n\rightarrow 0$ has to be taken at the end of the computations.
We have assumed that the disorder terms respect the isotropy (rotational invariance) in the 2d position space.

Using the ansatz of a replica-diagonal solution, and taking the limit $n\rightarrow 0$, we obtain a self-energy that is diagonal in the replica space, and can be written as
\begin{align}
 \Sigma^{2d}_{\text{dis}}(k_0,\mathbf k) 
& =  
2\,\mu^{\epsilon} \int\frac{d^{ 2-\epsilon}\mathbf q }{(2\,\pi)^{ 2-\epsilon}} 
\,G_0 (k_0,\mathbf q) = 
\frac{i \,k_0 \left(\frac{\mu } {| k_0 | }\right)^{\epsilon } } 
{\pi \, \epsilon }
\left( W_0 + W_1 +2\,W_2\right)+\mathcal{O}(\epsilon^0)\,.
\end{align}

Next let us consider the loop corrections to the disorder lines themselves.
A ZS diagram comes with a factor of $n$, which vanishes upon taking the replica limit $ n \rightarrow 0 $.
The contributions from the VC, BCS, and ZS$^{\prime}$ diagrams are shown in Tables \ref{vc2d} and \ref{bcs2d} respectively.

\begin{table}[h]
\begin{tabular}{|c|c|c|c|}
\hline
Coupling & $ W_{0}$ & $ W_{1} $ & $ W_{2}$   
\tabularnewline \hline
$ W_{0}$ & \makecell{$ \delta W_0=-\frac{2 \,W_0^2 }
{\pi  \, \epsilon } $ }
& \makecell{$ \delta W_0=-\frac{2 \, W_0\, W_1  } {\pi \, \epsilon } $}  
&   \makecell{$ \delta W_0=- \frac{4 \,W_0\, W_2 }   {\pi \, \epsilon } $}
\tabularnewline \hline
$ W_{1}$ &
\makecell{$ \delta W_1 = \frac{2\,W_0 \,W_1 }
{\pi \, \epsilon }$}
& \makecell{$ \delta W_1 = \frac{2 \,W_1^2 } {\pi \, \epsilon } $}
& \makecell{$ \delta W_1=-\frac{4\, W_0 \,W_2 }{\pi \,\epsilon } $}  \tabularnewline \hline
$ W_{2}$ &  
\makecell{$ \delta W_2=- \frac{2 \, W_0 \,W_2 }{\pi \,\epsilon }$}
& \makecell{ $  \delta W_2=\frac{2 \, W_1 \,W_2 } {\pi \, \epsilon }$ }&
\makecell{$ 0$}   \tabularnewline \hline
\end{tabular}
\caption{Corrections to disorder couplings in 2d from the VC diagrams involving disorder only.
\label{vc2d}} 
\end{table}

\begin{table}[h]
\begin{tabular}{|c|c|c|c|}
\hline
Coupling & $ W_{0}$ & $ W_{1} $ & $ W_{2}$   
\tabularnewline \hline
$ W_{0}$ & \makecell{$ \delta W_1= -\frac{ 2\,W_0^2  }   { \pi \, \epsilon }$ }
& \makecell{$ 0 $}  
&   \makecell{$ \delta  W_0 = -\frac{ 4\,W_0 \,W_2 }
{ \pi \, \epsilon } $}
\tabularnewline \hline
$ W_{1}$ &
\makecell{$ - $}
& \makecell{$ \delta W_2=-\frac{  W_1^ 2 } { \pi \, \epsilon } $}
& \makecell{$ \delta  W_0 = \frac{ 2\,W_1\, W_2}
{\pi \, \epsilon } ,\,
\delta  W_1=-\frac{ 6\,W_1\, W_2} {\pi \, \epsilon }  $}  
\tabularnewline \hline
$ W_{2}$ &  
\makecell{$-$}
& \makecell{ $ - $ }&
\makecell{$ \delta  W_2 = -\frac{ 8\,W_2^2} 
{\pi \,  \epsilon } $}   \tabularnewline \hline
\end{tabular}
\caption{Corrections to disorder couplings in 2d from the BCS and ZS$^\prime$ diagrams involving disorder only. Only the upper triangular part is populated as the lower triangular part contains duplicate entries.
\label{bcs2d}} 
\end{table}

Lastly, we need to consider the loop-diagrams, each of which originates from a Coulomb vertex and a disorder vertex. The loop integrals for the ZS diagrams vanish.
The VC diagrams add a correction term of $ \frac{2 \, e^2 \left(W_0+W_1 + 2 \,W_2\right)} {\pi \, \epsilon } $ to the bare coupling of $ - e^2 $. The BCS and ZS$^\prime$ diagrams have no logarithmic divergence.

\subsection{3d case}

The minimal set of disorder, including a scalar vertex, is given by
\begin{align}
\mathcal{S}^{3d}_{\text{dis}} & = -  \sum \limits_{a,b=1}^n 
 \int d\tau\, d\tau' \,d^{3-\epsilon} \mathbf x \left(
{ \bar{\mathcal V}\, \mu^{\epsilon-1}}
+ \frac { {\mathcal V} \, \mu^{\epsilon}} {|\mathbf x|^2}  \right)
\left( \psi_a^{\dag} \, {\psi_a}\right)_{\mathbf x,\tau}
\left ( \psi_b^{\dag} \, \psi_b \right )_{\mathbf x,\tau'} ,
\end{align}
with the Fourier transformed expression
\begin{align}
\mathcal{S}^{3d}_{\text{dis}} & = 
 - \sum \limits_{a,b }  
\int \frac{dk_0\, dk_0' \,
d^{3-\epsilon} {\mathbf q}\, d^{3-\epsilon} {\mathbf k}\, d^{3-\epsilon} {\mathbf k'}}
{(2\pi)^{3 \left (3-\epsilon \right )+2}}\, 
\left \lbrace \tilde{\psi}_a^{\dag}  (k_0,\mathbf k)\,
\tilde{\psi}_a  (k_0,{\mathbf k}+\mathbf q) \right \rbrace
\left \lbrace \tilde{\psi}^{\dag}_b ( k_0' ,{\mathbf k}')\, 
\tilde{\psi}_b (k_0',{\mathbf k}'-\mathbf q) \right \rbrace
\left( \bar{\mathcal V} \,\mu^{\epsilon-1}
+ \frac{ {\mathcal V} \,\mu^{\epsilon} }
{|\mathbf q|} \right ).
\end{align}

The fermion self-energy correction from disorder takes the form:
\begin{align}
 \Sigma^{3d}_{dis}(k_0,\mathbf k)
& =  
\frac{i \,k_0 \left(1+\zeta ^2-3 \, \zeta \sum \limits_{j=1}^3 \Gamma_{jj} \right)
\mathcal{V} \left(\frac{\mu } {| k_0 | }\right)^{\epsilon }}
{2 \, \pi  \left(1-\zeta ^2 \right)^2 \epsilon } \,.
\end{align}
The presence of the term $\sum \limits_{j=1}^3 \Gamma_{jj}$ indicates that a mass term is generated, which however does not fully gap out all the bands (as it does not anticommute with all terms of the Hamiltonian). We will ignore this term in our RG framework.

The disorder-disorder ZS contribution vanishes as we take the number of replicas $n \rightarrow 0$ limit. For the disorder-only VC diagrams, we get the correction
$ \frac{2\, \left( 1 +\zeta ^2 \right)}
{ \pi ^2 \left( 1 - \zeta ^2 \right)^2 \epsilon }
\,\mathcal{V} \,\bar{\mathcal{V}} $ and
$ \frac{2\, \left( 1 +\zeta ^2 \right)}
{ \pi ^2 \left( 1 - \zeta ^2 \right)^2  \epsilon }
\,\mathcal{V}^2 \,,$
to the bare couplings $-\bar{\mathcal{V}}$ and $ -{\mathcal{V}}$, respectively.
The disorder-disorder BCS and ZS$^\prime$ diagrams have no logarithmic divergences.

Lastly, we consider the mixed Coulomb-disorder diagrams. The ZS ones contribute to 
$ \delta {\bar{\mathcal V}} = 
  \frac{ e^2 \,\bar{\mathcal V}}
{3 \,\pi ^2\,  \epsilon } $ and
$ \delta {\mathcal V} = 
  \frac{  e^2 \,\mathcal V }
{3 \,\pi ^2\,  \epsilon }\,.$
The VC diagrams shift the Coulomb coupling by $
\delta (e^2) = -  \frac{ 2 \left(1+ \zeta ^2 \right)  }
{\pi ^2 \left(1-\zeta ^2\right)^2  \epsilon }
\,e^2\, \mathcal{V} $.
The BCS and ZS$^\prime$ diagrams have no logarithmic divergence.

\section{RG flow analysis}
\label{secrg}

Using the results from the one-loop calculations, we now compute the RG flow equations to determine the fixed points. We analyze the equations in two separate subsections for the 2d and 3d systems, respectively.

\subsection{2d case}

To the counterterm action in Eq.~\eqref{counter1}, we now add
the one required for the disorder. This is represented by 
\begin{align}
\mathcal{S}^{2d,c}_{\text{dis}} & = - \mu^{\epsilon} \sum \limits_{a,b=1}^n 
 \int d {k_0} \, d {p_0} \,
\left(\prod \limits_{ \eta =1}^4 \int  d^{2-\epsilon} \vec{k}_{\eta} \right)
 \frac{\delta^{2-\epsilon}
\left( \vec k_{1} + \vec k_{3}-\vec k_{2}-\vec k_{4} \right) 
}
{\left(2\,\pi\right)^{2+3\left(2-\epsilon \right) }}   
\Big[\,A_5\, {W_0} \left \lbrace 
\tilde \psi_a^{\dag} ( {k_0},\mathbf k_{1}) \, 
 {\tilde \psi_a} ( {k_0},\mathbf k_{2})\right \rbrace
\left \lbrace \tilde \psi_b^{\dag} ( {p_0},\mathbf k_{3})\, 
\tilde \psi_b ( {p_0},\mathbf k_{4})\right \rbrace
\nn & \hspace{ 6 cm} 
+ A_6\, {W_1} \left \lbrace  \tilde \psi_a^{\dag}( {k_0},\mathbf k_{1})
 \,\Gamma_{30}\, 
{\tilde \psi_a} ( {k_0},\mathbf k_{2})\right \rbrace
\left \lbrace \tilde \psi_b^{\dag} ( {p_0},\mathbf k_{3})\,\Gamma_{30}\, 
\tilde \psi_b ( {p_0},\mathbf k_{4})
\right \rbrace
\nn & \hspace{ 6 cm}
+ A_7\, {W_2} \sum \limits_{j=1,2}
\left \lbrace  \tilde \psi_a^{\dag}( {k_0},\mathbf k_{1})
 \,\Gamma_{j0}\, 
{\tilde \psi_a} ( {k_0},\mathbf k_{2})\right \rbrace
\left \lbrace \tilde \psi_b^{\dag} ( {p_0},\mathbf k_{3})\,\Gamma_{j0}\, 
\tilde \psi_b ( {p_0},\mathbf k_{4})
\right \rbrace \Big ]\, ,\nn
& A_{n} \equiv \,Z_n-1 = 
\sum_{\lambda=1}^\infty \frac{Z_{ n,\lambda}}
{\epsilon^\lambda}  \text{ with }  n=5,6,7  \,,
\end{align}
where
\begin{align}
{W_0}_B = Z_5\, Z_2^{\epsilon -2} \,W_0 \, \mu ^{\epsilon}\,,\quad
{W_1}_B = Z_6\, Z_2^{\epsilon -2} \,W_1 \, \mu ^{\epsilon}\,,\quad
{W_2}_B = Z_7\, Z_2^{\epsilon -2} \,W_2 \, \mu ^{\epsilon}\,.
\end{align}
We then get the corresponding renormalized action as
\begin{align}
\mathcal{S}^{2d,ren}_{\text{dis}} & = -  \sum \limits_{a,b=1}^n 
 \int d k_{0_B}\, d p_{0_B} \,
\left(\prod \limits_{ \eta =1}^4 \int  d^{2-\epsilon} \vec{k}_{\eta_B} \right)
\frac{\delta^{2-\epsilon}
\left( \vec k_{1_B} + \vec k_{3_B}-\vec k_{2_B}-\vec k_{4_B} \right) 
}
{\left(2\,\pi\right)^{2+3\left(2-\epsilon \right) }}   
\nn & \hspace{ 2 cm} \times
\Big[\,{W_0}_B \left \lbrace 
\tilde \psi_a^{\dag} ( {k_0}_B,\mathbf k_{1_B}) \, 
 {\tilde \psi_a} ( {k_0}_B,\mathbf k_{2_B})\right \rbrace
\left \lbrace \tilde \psi_b^{\dag} ( {p_0}_B,\mathbf k_{3_B})\, 
\tilde \psi_b ( {p_0}_B,\mathbf k_{4_B})\right \rbrace
\nn & \hspace{ 2.5 cm} + {W_1}_B \left \lbrace  \tilde \psi_a^{\dag}( {k_0}_B,\mathbf k_{1_B})
 \,\Gamma_{30}\, 
{\tilde \psi_a} ( {k_0}_B,\mathbf k_{2_B})\right \rbrace
\left \lbrace \tilde \psi_b^{\dag} ( {p_0}_B,\mathbf k_{3_B})\,\Gamma_{30}\, 
\tilde \psi_b ( {p_0}_B,\mathbf k_{4_B})
\right \rbrace 
\nn & \hspace{ 2.5 cm}+ {W_2}_B \sum \limits_{j=1,2}
\left \lbrace \tilde \psi_a^{\dag} ( {k_0}_B,\mathbf k_{1_B})
\,\Gamma_{j0}\, {\tilde \psi_a} ( {k_0}_B,\mathbf k_{2_B})\right \rbrace
\left \lbrace \tilde \psi_b^{\dag}  ( {p_0}_B,\mathbf k_{3_B})
\,\Gamma_{j0}\, 
\tilde \psi_b  ( {p_0}_B,\mathbf k_{4_B})
\right \rbrace \, \Big ]\, .
\end{align}

Since we have set $\zeta = 0$, the $Z_3$ is no longer in consideration.
To one-loop order, we now have
\begin{align}
 \label{Zdis2d} 
& Z_{1,1}=-\frac{W_0+W_1+2 W_2}{\pi }\,,\quad 
Z_{2,1}=-\frac{e^2}{4 \,\pi }\,,
\quad Z_{4,1}=-\frac{2 (W_0+W_1+2 W_2)}{\pi }\,,
\nn
&  Z_{5,1}=\frac{-2 \,W_0^2-2 W_0 \,W_1-4 W_0 \,W_2-4 \,W_0\, W_2
+2 \,W_1 \,W_2} {\pi \, W_0}\,,\quad 
Z_{6,1}=\frac{-2\, W_0^2+2 \,W_0 \,W_1
-4 \,W_0\, W_2+2\, W_1^2-6\, W_1 \,W_2}
{\pi \, W_1}\,,\nn &
Z_{7,1}=\frac{-2 \,W_0 \,W_2-W_1^2+2 \,W_1\, W_2-8 W_2^2}{\pi  W_2}\,.
\end{align}
We have used the results from Tables \ref{vc2d} and \ref{bcs2d}.

The beta functions are calculated in the same way as in the clean case. We get:
\begin{align}
& \frac{ de^2} {dl} = e^2 
\left[ \epsilon + 
\frac{ -e^2  +4 \left (W_0+W_1+2\, W_2 \right ) }
{4 \, \pi } \right], \quad
\frac{ dW_0} {dl} = \epsilon\,W_0 
+ \frac{4  \left[ W_0^2+  W_0 \left (W_1+4 \, W_2\right )
-W_1 \,W_2 \right ]- e^2\,  W_0}
{2 \,\pi} \,,\nn
& \frac{ dW_1 } {dl} = \epsilon\,W_1 
+ \frac{4  \left(W_0^2-W_0 \, W_1+2 \,W_0 \,W_2-W_1^2+3\, W_1 \,W_2\right)
- e^2\, W_1} {2 \,\pi}    \,, \nn &
\frac{ dW_2} {dl} = \epsilon\,W_2
+ \frac{2  \left[2 \,W_2 \left (W_0+4 W_2 \right )+W_1^2-2 \,W_1 \,W_2\right ]
- e^2\, W_2}
{2\, \pi} \,.
\end{align}

\begin{figure}[]
\begin{center}
\subfigure[\label{allt}]{\includegraphics[width=0.32 \textwidth]{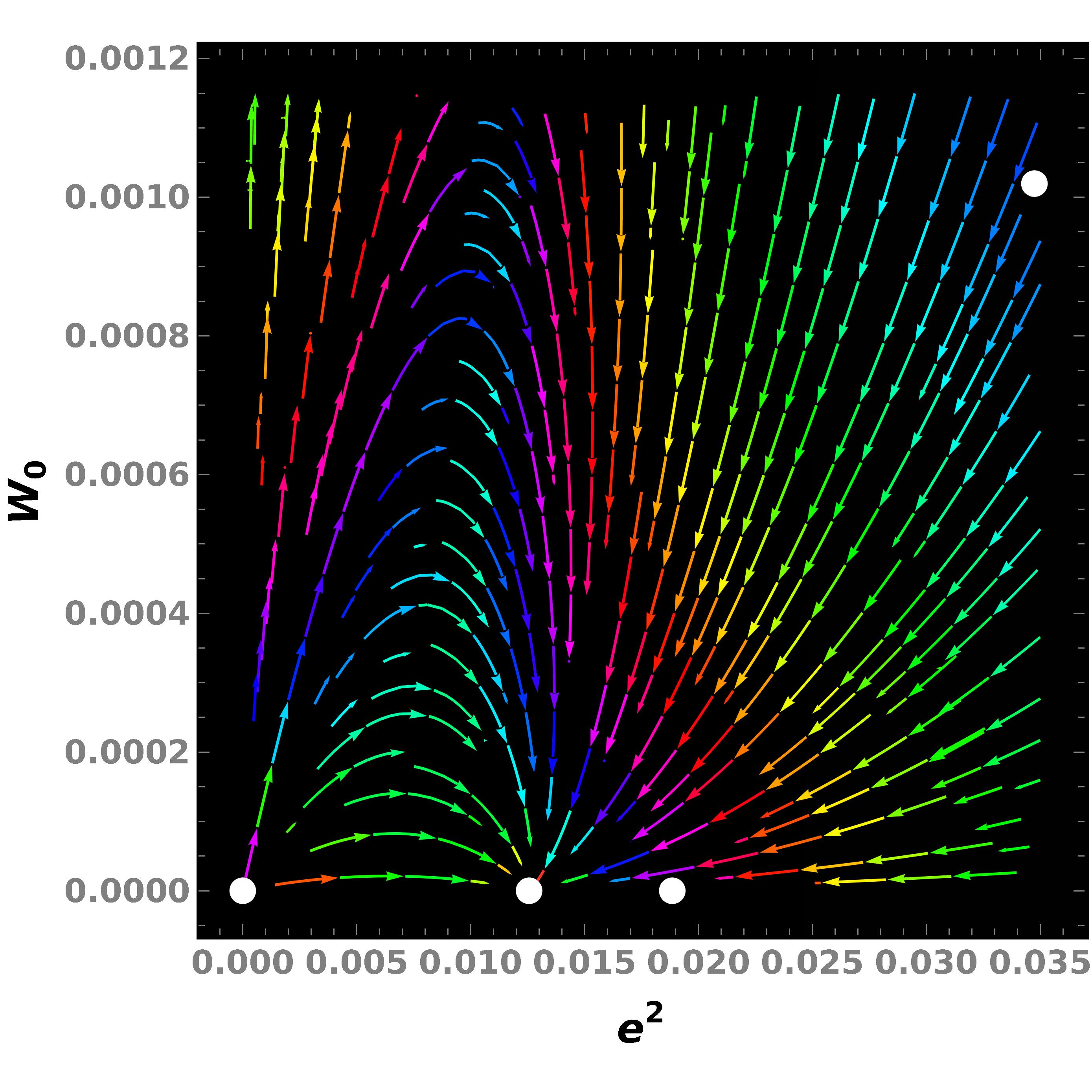}} \,
\subfigure[\label{noise1}]{\includegraphics[width=0.32\textwidth]{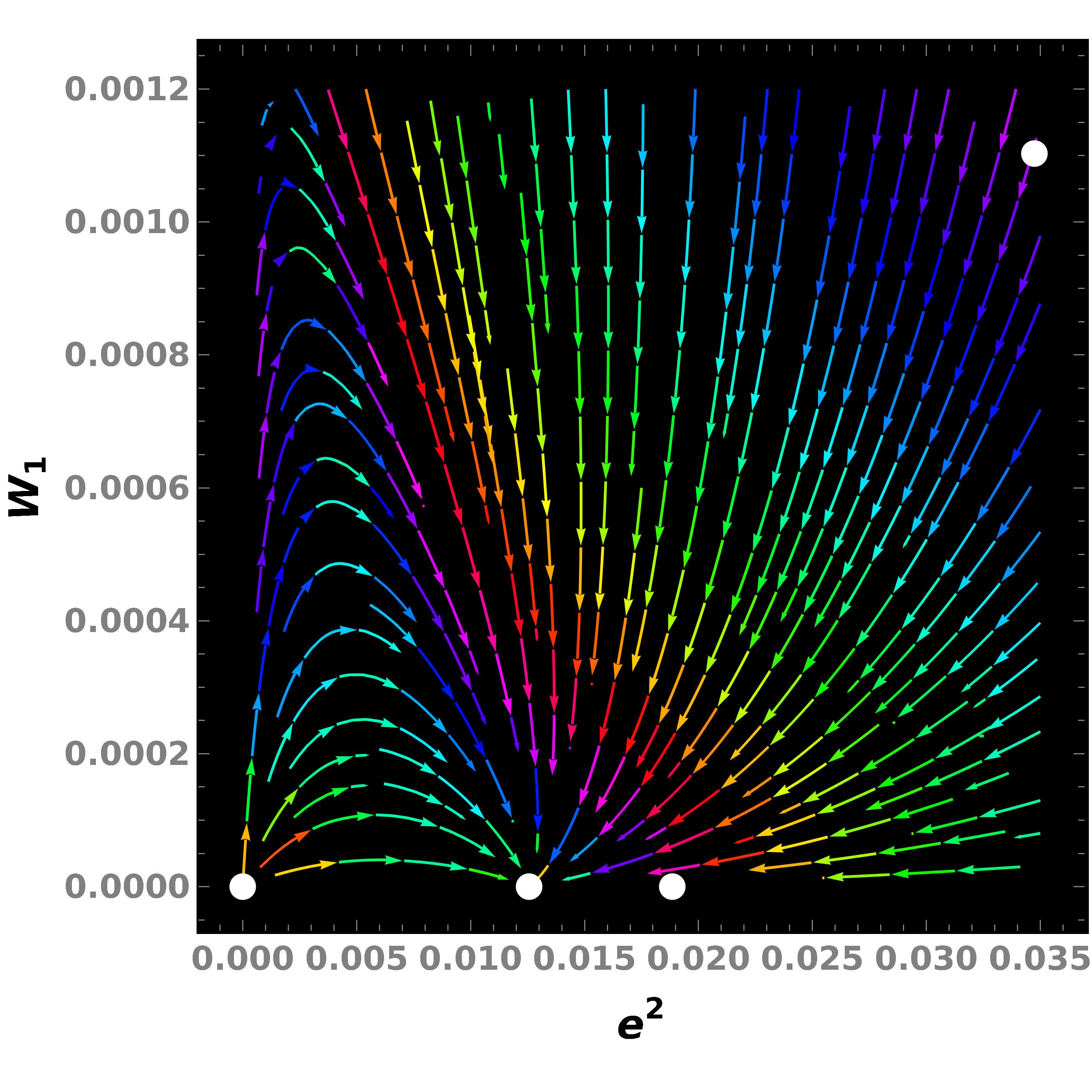}} \,
\subfigure[\label{derinoise}]{\includegraphics[width=0.32 \textwidth]{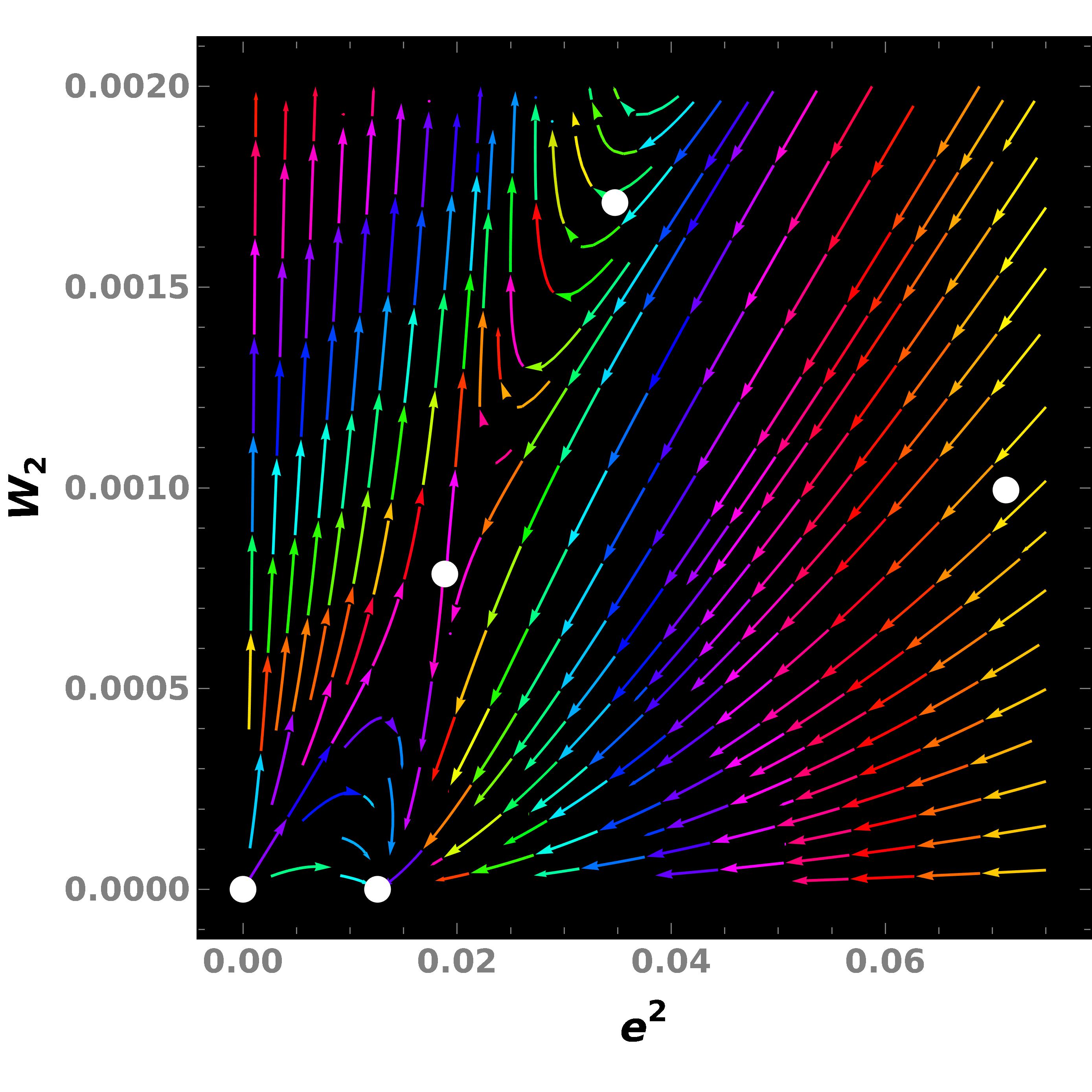}}
\caption{\label{figrg1}
2d case: Panels (a), (b), and (c) show the RG flows and fixed points (represented by white discs) in the $ W_1=W_2=0$, $ W_0=W_2=0$, and $ W_0=W_1=0$ planes, respectively. We have set $\epsilon=10^{-3} $ for all the plots.}
\end{center}	
\end{figure}

The fixed points are given by the zeros of the above derivatives of the four coupling constants. We get several fixed points, where the values of the coupling constants $\left \lbrace {e^2}, W_0, W_1, W_2 \right \rbrace$ are given by:
\begin{align}
\left \lbrace  0,0,0,0 \right \rbrace ,\quad
\left \lbrace 4\, \pi \, \epsilon,0,0,0  \right \rbrace ,\quad
\left \lbrace  6 \,\pi \, \epsilon,0,0, \frac{\pi \, \epsilon }{4} \right \rbrace ,\quad
\left \lbrace 34.74 \, \epsilon ,
 1.02 \,\epsilon , 1.10 \,\epsilon ,
 1.71 \,\epsilon  \right \rbrace ,\quad
\left \lbrace  71.27\, \epsilon ,  8.93\, \epsilon ,
 3.76\, \epsilon ,  0.99 \, \epsilon \right \rbrace .
\end{align}
Expanding the coupling constants about each fixed point, we determine the stability
matrix. Only for the fixed point $\left \lbrace 4\, \pi \, \epsilon,0,0,0  \right \rbrace $, all eigenvalues of the stability matrix are negative, implying that it is the stable fixed point. The RG flows in three different planes are shown in Fig.~\ref{figrg1}. Hence, the marginal Fermi liquid fixed point survives in the presence of disorder.

\subsection{3d case}

To the counterterm action in Eq.~\eqref{counter1}, we now add
the one required for the disorder. This is represented by 
\begin{align}
\mathcal{S}^{3d,c}_{\text{dis}} & = -  \sum \limits_{a,b=1}^n
 \int d {k_0} \, d {p_0} \,
\left(\prod \limits_{ \eta =1}^4 \int  d^{3-\epsilon} \vec{k}_{\eta} \right)
\frac{\delta^{ 3-\epsilon}
\left( \vec k_{1} + \vec k_{3}-\vec k_{2}-\vec k_{4} \right)
}
{\left(2\,\pi\right)^{2+3\left(2-\epsilon \right) }}
\left( A_5\, \bar{\mathcal V}\,\mu^{\epsilon-1} 
+ \frac{ A_6 \,{\mathcal V} \,\mu^{\epsilon} }
{|\mathbf q|} \right )
\nn & \hspace{ 5 cm }
\times \left \lbrace
\tilde \psi_a^{\dag} ( {k_0},\mathbf k_{1}) \,
 {\tilde \psi_a} ( {k_0},\mathbf k_{2})\right \rbrace
\left \lbrace \tilde \psi_b^{\dag} ( {p_0},\mathbf k_{3})\,
\tilde \psi_b ( {p_0},\mathbf k_{4})\right \rbrace,\nn
& A_{n} \equiv \,Z_n-1 =
\sum_{\lambda=1}^\infty \frac{Z_{ n,\lambda}}
{\epsilon^\lambda}  \text{ with }  n=5,6  \,,
\end{align}
where
\begin{align}
{\bar{\mathcal V}}_B = Z_5\, Z_2^{\epsilon -3} \,{\bar{\mathcal V}} 
\, \mu ^{\epsilon-1}\,,\quad
{\mathcal V}_B = Z_6\, Z_2^{\epsilon -2} \,{\mathcal V} 
\, \mu ^{\epsilon}\,.
\end{align}
We then get the corresponding renormalized action as
\begin{align}
\mathcal{S}^{3d,ren}_{\text{dis}} & = -  \sum \limits_{a,b=1}^n 
 \int d k_{0_B}\, d p_{0_B} \,
\left(\prod \limits_{ \eta =1}^4 \int  d^{3-\epsilon} \vec{k}_{\eta_B} \right)
\frac{\delta^{3-\epsilon}
\left( \vec k_{1_B} + \vec k_{3_B}-\vec k_{2_B}-\vec k_{4_B} \right) 
}
{\left(2\,\pi\right)^{2+3\left(3-\epsilon \right) }}  
\left(   \bar{\mathcal V}_B 
+ \frac{ {\mathcal V}_B} {|\mathbf q_B|} \right ) 
\nn & \hspace{ 5 cm} \times
 \left \lbrace 
\tilde \psi_a^{\dag} ( {k_0}_B,\mathbf k_{1_B}) \, 
 {\tilde \psi_a} ( {k_0}_B,\mathbf k_{2_B})\right \rbrace
\left \lbrace \tilde \psi_b^{\dag} ( {p_0}_B,\mathbf k_{3_B})\, 
\tilde \psi_b ( {p_0}_B,\mathbf k_{4_B})\right \rbrace .
\end{align}

To one-loop order, we have
\begin{align}
 \label{Zdis2d} 
Z_{1,1} &= - \frac{\left(1+\zeta ^2 \right) \mathcal{V} }
{2 \, \pi \left(1-\zeta ^2 \right)^2 } \,,\quad
 Z_{2,1}  =  -\frac{e^2} { 3\,\pi^2 } \,,
\quad Z_{3,1} =0\,,\quad
Z_{4,1}  = \frac{e^2} { 3\,\pi^2}
-  \frac{ 2 \left(1+ \zeta ^2 \right)   \mathcal{V} }
{\pi^2 \left(1-\zeta ^2\right)^2  }\,,
\nn Z_{5,1} & = -\frac{2\, \left( 1 +\zeta ^2 \right)
\mathcal{V} }
{ \pi^2  \left( 1 - \zeta ^2 \right)^2 } 
+ \frac{ e^2  }
{3 \,\pi ^2 }\,,\quad
Z_{6,1} = - \frac{2\, \left( 1 +\zeta ^2 \right)
\mathcal{V} } { \pi^2 \left( 1 - \zeta ^2 \right)^2 } 
+  \frac{  e^2 } {3 \,\pi ^2} \,.
\end{align}

\begin{figure}[]
\begin{center}
\subfigure[\label{allt}]{\includegraphics[width=0.32 \textwidth]{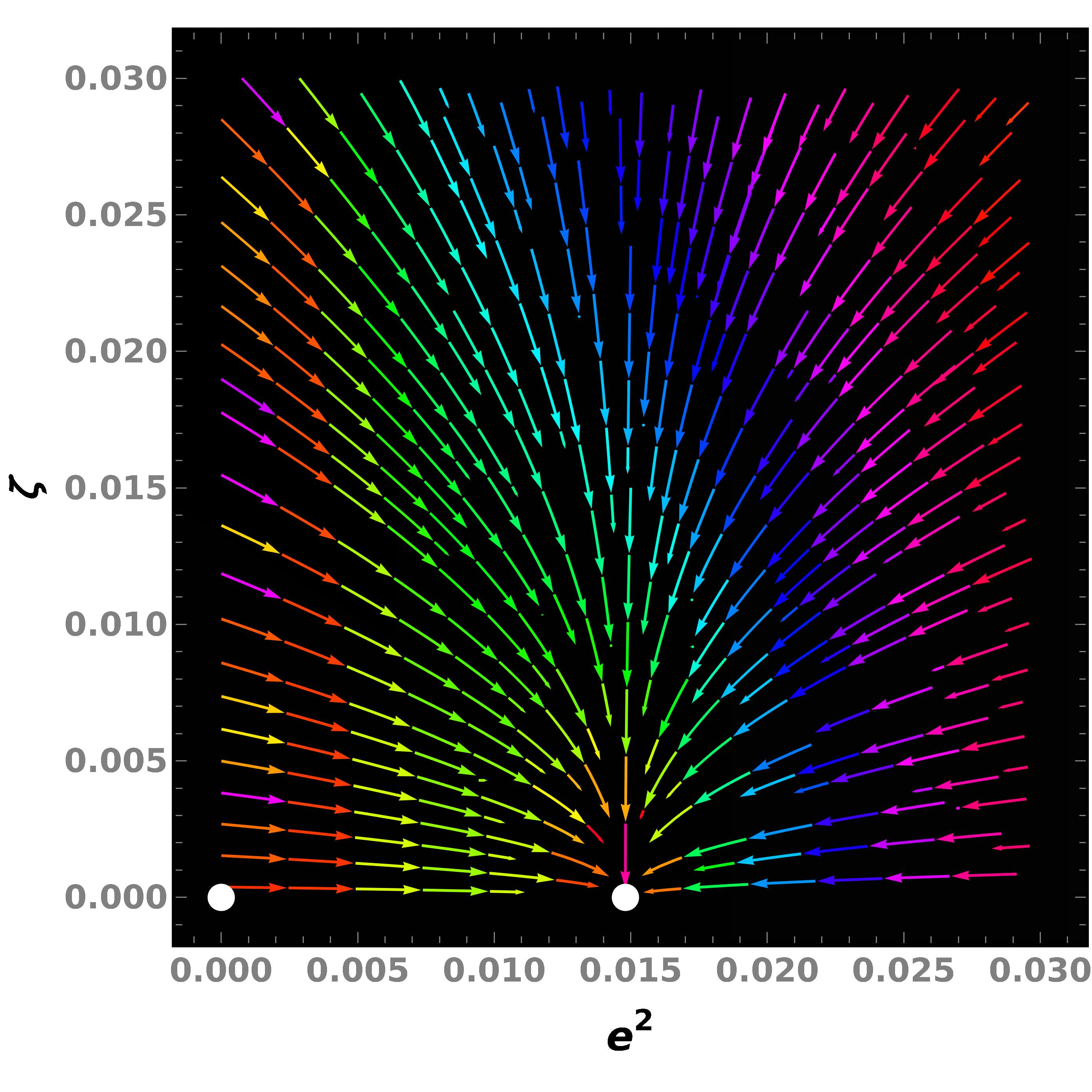}} \,
\subfigure[\label{noise1}]{\includegraphics[width=0.32\textwidth]{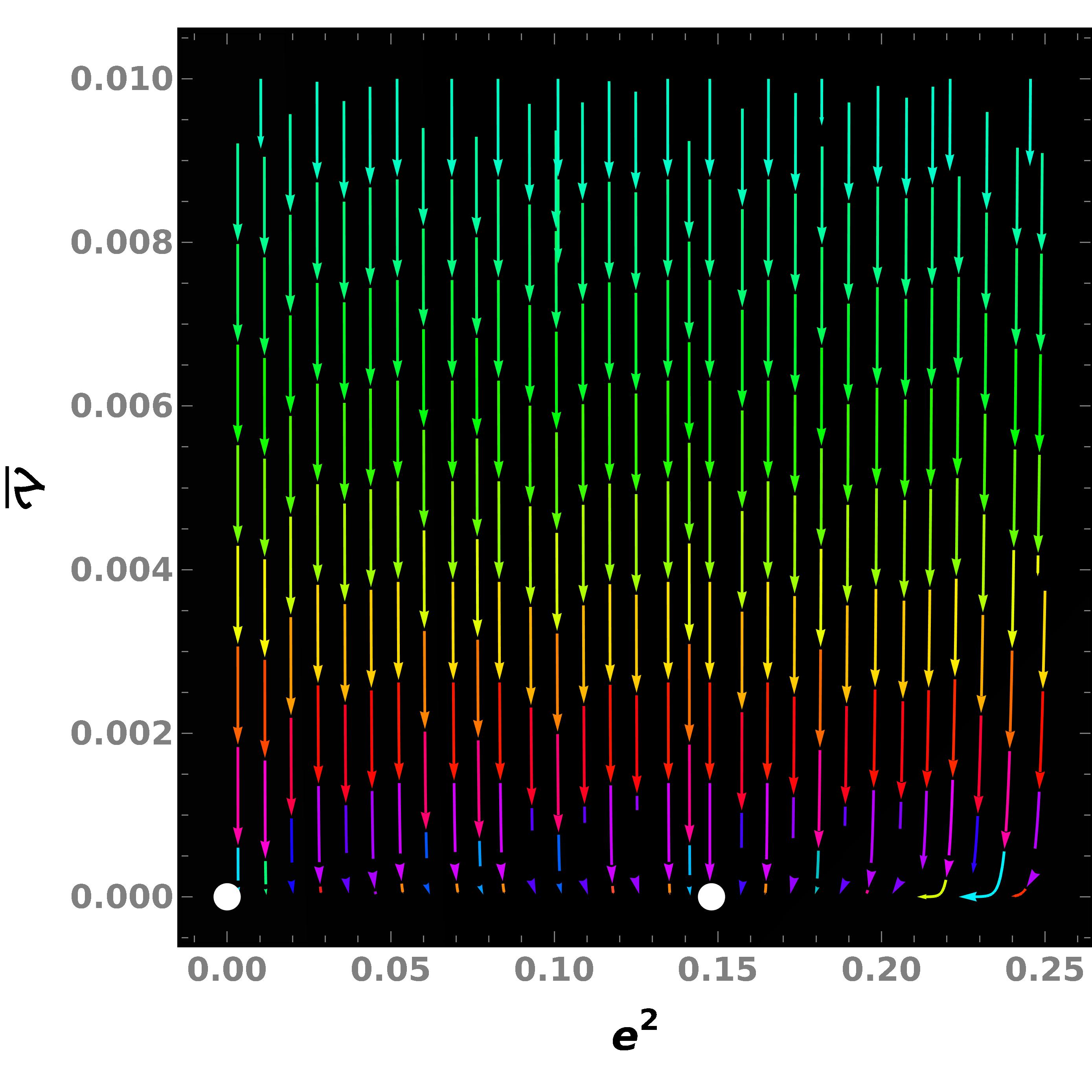}} \,
\subfigure[\label{derinoise}]{\includegraphics[width=0.32 \textwidth]{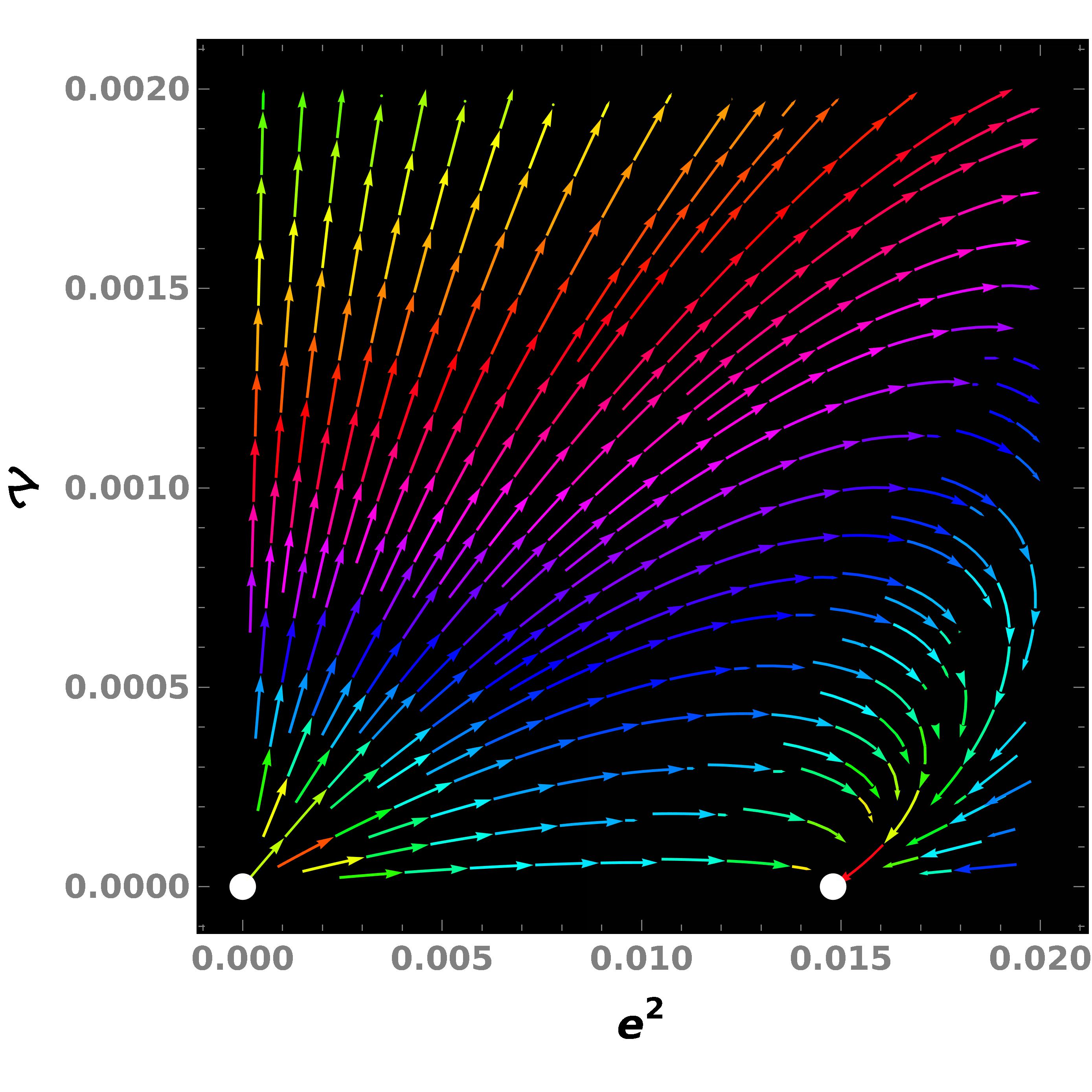}}
\caption{\label{figrg2}
3d case: Panels (a), (b), and (c) show the RG flows and fixed points (represented by white discs) in the $ \bar{\mathcal V}={\mathcal V}=0$, $ \zeta={\mathcal V}=0$, and $ \zeta=\bar{\mathcal V}=0$ planes, respectively.
We have set $\epsilon=10^{-3} $ for (a) and (c), and $\epsilon=10^{-2} $ for (b).}
\end{center}	
\end{figure}

The full set of beta functions is given by:
\begin{align}
& \frac{ de^2} {dl} = e^2 
\left[ \epsilon - 
\frac{ 4 \,e^2 +\frac{3 (\pi -4) 
\left( 1 +\zeta ^2 \right) \mathcal V}
{\left(1-\zeta ^2\right)^2} } {6 \, \pi ^2  }
\right], \quad
\frac{ d\zeta} {dl} = -\frac{e^2 \,\zeta } {3 \,\pi ^2} \,,\nn
& \frac{ d \bar{\mathcal V}} {dl} = \bar{\mathcal V} 
\left [ \epsilon -1
+ \frac{2 \left \lbrace \frac{3 \,\pi  
\left(\zeta ^2+1\right) \mathcal{V}}{\left(1-\zeta ^2\right)^2}
-2 \,e^2\right  \rbrace }
{3 \,\pi ^2} \right],\quad
\frac{ d \mathcal V} {dl} = \mathcal V \left [ \epsilon
+ \frac{\frac{2 \,\pi  \left( 1 + \zeta ^2\right) \mathcal{V}}
{\left(1-\zeta ^2\right)^2}-e^2} {\pi ^2} \right] .
\end{align}
The fixed points of the beta functions are given by:
\begin{align}
\left \lbrace  0,0,0,0 \right \rbrace ,\quad
\left \lbrace \frac{3}{2} \,\pi ^2 \, \epsilon,0,0,0  \right \rbrace  .
\end{align}
Only for the non-Gaussian fixed point, all eigenvalues of the stability matrix are negative, implying that it is the stable fixed point. The RG flows in three different planes are shown in Fig.~\ref{figrg2}. Hence, the marginal Fermi liquid fixed point survives in the presence of disorder.

\section{Summary and Outlook}
\label{conclude}

We have considered effective low-energy Hamiltonians which give rise to pseudospin-3/2 quasiparticles with birefringent spectra, in both 2d and 3d.
First, we have computed the stable phases of these systems in the presence of Coulomb interactions by using the RG scheme. Although this question was considered earlier \cite{malcolm-bitan}, we have found it essential to repeat the derivations to correct some algebraic factors in the loop-calculations, and also to set up our minimal subtraction scheme. The results show that a marginal Fermi liquid emerges both in 2d and 3d, driven by the Coulomb interactions, which is a stable interacting fixed point of the RG equations. Interestingly, this is an example where a marginal Fermi liquid emerges even in 2d, where for other systems we usually find a non-Fermi liquid phase \cite{denis,ips-uv-ir1,ips-uv-ir2,ips-nfl-u1,ips-fflo}.

The focal point of our work is to analyze if this non-Gaussian fixed point survives the addition of disorder, which would be ubiquitous in realistic scenarios. We have treated disorder on an equal footing with the Coulomb terms, and have re-derived the RG equations, now in the presence of disorder vertices. The stable fixed points of the RG equations show that the marginal Fermi liquid is robust against disorder.

In future, it will be useful to compute the close-to-zero temperature transport properties, by using methods like Kubo formula \cite{ips-subir,ips-c2}, and the invariant measure approach (IMA) \cite{ips-klaus}.
The finite-temperature transport characteristics can be calculated by using the memory matrix formalism \cite{ips-freire1,ips-freire2}. In particular, since the marginal Fermi liquid phase is unaffected by disorder, there is no subtlety
(unlike in similar non-Fermi liquid phases in other semimetals, considered earlier \cite{ips-freire1,ips-freire2}) in applying the memory matrix formalism, where we usually couple the system with weak disorder in order to provide a relaxation mechanism for all the nearly-conserved operators. Another interesting direction will be to consider the effect of disorder in the presence of various order parameters / mass terms \cite{malcolm-bitan,bitan-vladimir,ips-qbt-sc}.

\section{Acknowledgments}
We thank Klaus Ziegler and Vladimir Juri\ifmmode \check{c}\else \v{c}\fi{}i\ifmmode \acute{c}\else \'{c}\fi{} for useful comments.

\appendix

\bibliography{biblio}

\begin{thebibliography}{27}%
\makeatletter
\providecommand \@ifxundefined [1]{%
 \@ifx{#1\undefined}
}%
\providecommand \@ifnum [1]{%
 \ifnum #1\expandafter \@firstoftwo
 \else \expandafter \@secondoftwo
 \fi
}%
\providecommand \@ifx [1]{%
 \ifx #1\expandafter \@firstoftwo
 \else \expandafter \@secondoftwo
 \fi
}%
\providecommand \natexlab [1]{#1}%
\providecommand \enquote  [1]{``#1''}%
\providecommand \bibnamefont  [1]{#1}%
\providecommand \bibfnamefont [1]{#1}%
\providecommand \citenamefont [1]{#1}%
\providecommand \href@noop [0]{\@secondoftwo}%
\providecommand \href [0]{\begingroup \@sanitize@url \@href}%
\providecommand \@href[1]{\@@startlink{#1}\@@href}%
\providecommand \@@href[1]{\endgroup#1\@@endlink}%
\providecommand \@sanitize@url [0]{\catcode `\\12\catcode `\$12\catcode
  `\&12\catcode `\#12\catcode `\^12\catcode `\_12\catcode `\%12\relax}%
\providecommand \@@startlink[1]{}%
\providecommand \@@endlink[0]{}%
\providecommand \url  [0]{\begingroup\@sanitize@url \@url }%
\providecommand \@url [1]{\endgroup\@href {#1}{\urlprefix }}%
\providecommand \urlprefix  [0]{URL }%
\providecommand \Eprint [0]{\href }%
\providecommand \doibase [0]{https://doi.org/}%
\providecommand \selectlanguage [0]{\@gobble}%
\providecommand \bibinfo  [0]{\@secondoftwo}%
\providecommand \bibfield  [0]{\@secondoftwo}%
\providecommand \translation [1]{[#1]}%
\providecommand \BibitemOpen [0]{}%
\providecommand \bibitemStop [0]{}%
\providecommand \bibitemNoStop [0]{.\EOS\space}%
\providecommand \EOS [0]{\spacefactor3000\relax}%
\providecommand \BibitemShut  [1]{\csname bibitem#1\endcsname}%
\let\auto@bib@innerbib\@empty
\bibitem [{\citenamefont {Kennett}\ \emph {et~al.}(2011)\citenamefont
  {Kennett}, \citenamefont {Komeilizadeh}, \citenamefont {Kaveh},\ and\
  \citenamefont {Smith}}]{malcolm}%
  \BibitemOpen
  \bibfield  {author} {\bibinfo {author} {\bibfnamefont {M.~P.}\ \bibnamefont
  {Kennett}}, \bibinfo {author} {\bibfnamefont {N.}~\bibnamefont
  {Komeilizadeh}}, \bibinfo {author} {\bibfnamefont {K.}~\bibnamefont
  {Kaveh}},\ and\ \bibinfo {author} {\bibfnamefont {P.~M.}\ \bibnamefont
  {Smith}},\ }\bibfield  {title} {\bibinfo {title} {Birefringent breakup of
  dirac fermions on a square optical lattice},\ }\href
  {https://doi.org/10.1103/PhysRevA.83.053636} {\bibfield  {journal} {\bibinfo
  {journal} {Phys. Rev. A}\ }\textbf {\bibinfo {volume} {83}},\ \bibinfo
  {pages} {053636} (\bibinfo {year} {2011})}\BibitemShut {NoStop}%
\bibitem [{\citenamefont {Roy}\ \emph {et~al.}(2012)\citenamefont {Roy},
  \citenamefont {Smith},\ and\ \citenamefont {Kennett}}]{prb.85.235119}%
  \BibitemOpen
  \bibfield  {author} {\bibinfo {author} {\bibfnamefont {B.}~\bibnamefont
  {Roy}}, \bibinfo {author} {\bibfnamefont {P.~M.}\ \bibnamefont {Smith}},\
  and\ \bibinfo {author} {\bibfnamefont {M.~P.}\ \bibnamefont {Kennett}},\
  }\bibfield  {title} {\bibinfo {title} {Asymmetric spatial structure of zero
  modes for birefringent dirac fermions},\ }\href
  {https://doi.org/10.1103/PhysRevB.85.235119} {\bibfield  {journal} {\bibinfo
  {journal} {Phys. Rev. B}\ }\textbf {\bibinfo {volume} {85}},\ \bibinfo
  {pages} {235119} (\bibinfo {year} {2012})}\BibitemShut {NoStop}%
\bibitem [{\citenamefont {Komeilizadeh}\ and\ \citenamefont
  {Kennett}(2014)}]{prb.90.045131}%
  \BibitemOpen
  \bibfield  {author} {\bibinfo {author} {\bibfnamefont {N.}~\bibnamefont
  {Komeilizadeh}}\ and\ \bibinfo {author} {\bibfnamefont {M.~P.}\ \bibnamefont
  {Kennett}},\ }\bibfield  {title} {\bibinfo {title} {Instabilities of a
  birefringent semimetal},\ }\href {https://doi.org/10.1103/PhysRevB.90.045131}
  {\bibfield  {journal} {\bibinfo  {journal} {Phys. Rev. B}\ }\textbf {\bibinfo
  {volume} {90}},\ \bibinfo {pages} {045131} (\bibinfo {year}
  {2014})}\BibitemShut {NoStop}%
\bibitem [{\citenamefont {D\'ora}\ \emph {et~al.}(2011)\citenamefont {D\'ora},
  \citenamefont {Kailasvuori},\ and\ \citenamefont {Moessner}}]{prb84.195422}%
  \BibitemOpen
  \bibfield  {author} {\bibinfo {author} {\bibfnamefont {B.}~\bibnamefont
  {D\'ora}}, \bibinfo {author} {\bibfnamefont {J.}~\bibnamefont
  {Kailasvuori}},\ and\ \bibinfo {author} {\bibfnamefont {R.}~\bibnamefont
  {Moessner}},\ }\bibfield  {title} {\bibinfo {title} {Lattice generalization
  of the dirac equation to general spin and the role of the flat band},\ }\href
  {https://doi.org/10.1103/PhysRevB.84.195422} {\bibfield  {journal} {\bibinfo
  {journal} {Phys. Rev. B}\ }\textbf {\bibinfo {volume} {84}},\ \bibinfo
  {pages} {195422} (\bibinfo {year} {2011})}\BibitemShut {NoStop}%
\bibitem [{\citenamefont {Watanabe}\ \emph {et~al.}(2011)\citenamefont
  {Watanabe}, \citenamefont {Hatsugai},\ and\ \citenamefont
  {Aoki}}]{watanabe_2011}%
  \BibitemOpen
  \bibfield  {author} {\bibinfo {author} {\bibfnamefont {H.}~\bibnamefont
  {Watanabe}}, \bibinfo {author} {\bibfnamefont {Y.}~\bibnamefont {Hatsugai}},\
  and\ \bibinfo {author} {\bibfnamefont {H.}~\bibnamefont {Aoki}},\ }\bibfield
  {title} {\bibinfo {title} {Manipulation of the dirac cones and the anomaly in
  the graphene related quantum hall effect},\ }\href
  {https://doi.org/10.1088/1742-6596/334/1/012044} {\bibfield  {journal}
  {\bibinfo  {journal} {Journal of Physics: Conference Series}\ }\textbf
  {\bibinfo {volume} {334}},\ \bibinfo {pages} {012044} (\bibinfo {year}
  {2011})}\BibitemShut {NoStop}%
\bibitem [{\citenamefont {Lan}\ \emph {et~al.}(2011{\natexlab{a}})\citenamefont
  {Lan}, \citenamefont {Goldman}, \citenamefont {Bermudez}, \citenamefont
  {Lu},\ and\ \citenamefont {\"Ohberg}}]{prb.84.165115}%
  \BibitemOpen
  \bibfield  {author} {\bibinfo {author} {\bibfnamefont {Z.}~\bibnamefont
  {Lan}}, \bibinfo {author} {\bibfnamefont {N.}~\bibnamefont {Goldman}},
  \bibinfo {author} {\bibfnamefont {A.}~\bibnamefont {Bermudez}}, \bibinfo
  {author} {\bibfnamefont {W.}~\bibnamefont {Lu}},\ and\ \bibinfo {author}
  {\bibfnamefont {P.}~\bibnamefont {\"Ohberg}},\ }\bibfield  {title} {\bibinfo
  {title} {Dirac-weyl fermions with arbitrary spin in two-dimensional optical
  superlattices},\ }\href {https://doi.org/10.1103/PhysRevB.84.165115}
  {\bibfield  {journal} {\bibinfo  {journal} {Phys. Rev. B}\ }\textbf {\bibinfo
  {volume} {84}},\ \bibinfo {pages} {165115} (\bibinfo {year}
  {2011}{\natexlab{a}})}\BibitemShut {NoStop}%
\bibitem [{\citenamefont {Lan}\ \emph {et~al.}(2011{\natexlab{b}})\citenamefont
  {Lan}, \citenamefont {Celi}, \citenamefont {Lu}, \citenamefont {\"Ohberg},\
  and\ \citenamefont {Lewenstein}}]{prl.107.253001}%
  \BibitemOpen
  \bibfield  {author} {\bibinfo {author} {\bibfnamefont {Z.}~\bibnamefont
  {Lan}}, \bibinfo {author} {\bibfnamefont {A.}~\bibnamefont {Celi}}, \bibinfo
  {author} {\bibfnamefont {W.}~\bibnamefont {Lu}}, \bibinfo {author}
  {\bibfnamefont {P.}~\bibnamefont {\"Ohberg}},\ and\ \bibinfo {author}
  {\bibfnamefont {M.}~\bibnamefont {Lewenstein}},\ }\bibfield  {title}
  {\bibinfo {title} {Tunable multiple layered dirac cones in optical
  lattices},\ }\href {https://doi.org/10.1103/PhysRevLett.107.253001}
  {\bibfield  {journal} {\bibinfo  {journal} {Phys. Rev. Lett.}\ }\textbf
  {\bibinfo {volume} {107}},\ \bibinfo {pages} {253001} (\bibinfo {year}
  {2011}{\natexlab{b}})}\BibitemShut {NoStop}%
\bibitem [{\citenamefont {Bradlyn}\ \emph {et~al.}(2016)\citenamefont
  {Bradlyn}, \citenamefont {Cano}, \citenamefont {Wang}, \citenamefont
  {Vergniory}, \citenamefont {Felser}, \citenamefont {Cava},\ and\
  \citenamefont {Bernevig}}]{bradlyn}%
  \BibitemOpen
  \bibfield  {author} {\bibinfo {author} {\bibfnamefont {B.}~\bibnamefont
  {Bradlyn}}, \bibinfo {author} {\bibfnamefont {J.}~\bibnamefont {Cano}},
  \bibinfo {author} {\bibfnamefont {Z.}~\bibnamefont {Wang}}, \bibinfo {author}
  {\bibfnamefont {M.~G.}\ \bibnamefont {Vergniory}}, \bibinfo {author}
  {\bibfnamefont {C.}~\bibnamefont {Felser}}, \bibinfo {author} {\bibfnamefont
  {R.~J.}\ \bibnamefont {Cava}},\ and\ \bibinfo {author} {\bibfnamefont
  {B.~A.}\ \bibnamefont {Bernevig}},\ }\bibfield  {title} {\bibinfo {title}
  {Beyond dirac and weyl fermions: Unconventional quasiparticles in
  conventional crystals},\ }\href {https://doi.org/10.1126/science.aaf5037}
  {\bibfield  {journal} {\bibinfo  {journal} {Science}\ }\textbf {\bibinfo
  {volume} {353}},\ \bibinfo {pages} {aaf5037} (\bibinfo {year}
  {2016})}\BibitemShut {NoStop}%
\bibitem [{\citenamefont {Ezawa}(2016)}]{prb.94.195205}%
  \BibitemOpen
  \bibfield  {author} {\bibinfo {author} {\bibfnamefont {M.}~\bibnamefont
  {Ezawa}},\ }\bibfield  {title} {\bibinfo {title} {Pseudospin-$\frac{3}{2}$
  fermions, type-ii weyl semimetals, and critical weyl semimetals in tricolor
  cubic lattices},\ }\href {https://doi.org/10.1103/PhysRevB.94.195205}
  {\bibfield  {journal} {\bibinfo  {journal} {Phys. Rev. B}\ }\textbf {\bibinfo
  {volume} {94}},\ \bibinfo {pages} {195205} (\bibinfo {year}
  {2016})}\BibitemShut {NoStop}%
\bibitem [{\citenamefont {Hsieh}\ \emph {et~al.}(2014)\citenamefont {Hsieh},
  \citenamefont {Liu},\ and\ \citenamefont {Fu}}]{PhysRevB.90.081112}%
  \BibitemOpen
  \bibfield  {author} {\bibinfo {author} {\bibfnamefont {T.~H.}\ \bibnamefont
  {Hsieh}}, \bibinfo {author} {\bibfnamefont {J.}~\bibnamefont {Liu}},\ and\
  \bibinfo {author} {\bibfnamefont {L.}~\bibnamefont {Fu}},\ }\bibfield
  {title} {\bibinfo {title} {Topological crystalline insulators and dirac
  octets in antiperovskites},\ }\href
  {https://doi.org/10.1103/PhysRevB.90.081112} {\bibfield  {journal} {\bibinfo
  {journal} {Phys. Rev. B}\ }\textbf {\bibinfo {volume} {90}},\ \bibinfo
  {pages} {081112} (\bibinfo {year} {2014})}\BibitemShut {NoStop}%
\bibitem [{\citenamefont {Chen}\ \emph {et~al.}(2017)\citenamefont {Chen},
  \citenamefont {Wang}, \citenamefont {Liu}, \citenamefont {Yu}, \citenamefont
  {Sheng}, \citenamefont {Chen},\ and\ \citenamefont
  {Yang}}]{PhysRevMaterials.1.044201}%
  \BibitemOpen
  \bibfield  {author} {\bibinfo {author} {\bibfnamefont {C.}~\bibnamefont
  {Chen}}, \bibinfo {author} {\bibfnamefont {S.-S.}\ \bibnamefont {Wang}},
  \bibinfo {author} {\bibfnamefont {L.}~\bibnamefont {Liu}}, \bibinfo {author}
  {\bibfnamefont {Z.-M.}\ \bibnamefont {Yu}}, \bibinfo {author} {\bibfnamefont
  {X.-L.}\ \bibnamefont {Sheng}}, \bibinfo {author} {\bibfnamefont
  {Z.}~\bibnamefont {Chen}},\ and\ \bibinfo {author} {\bibfnamefont {S.~A.}\
  \bibnamefont {Yang}},\ }\bibfield  {title} {\bibinfo {title} {Ternary
  wurtzite caagbi materials family: A playground for essential and accidental,
  type-i and type-ii dirac fermions},\ }\href
  {https://doi.org/10.1103/PhysRevMaterials.1.044201} {\bibfield  {journal}
  {\bibinfo  {journal} {Phys. Rev. Materials}\ }\textbf {\bibinfo {volume}
  {1}},\ \bibinfo {pages} {044201} (\bibinfo {year} {2017})}\BibitemShut
  {NoStop}%
\bibitem [{\citenamefont {Roy}\ \emph {et~al.}(2018)\citenamefont {Roy},
  \citenamefont {Kennett}, \citenamefont {Yang},\ and\ \citenamefont
  {Juri\ifmmode \check{c}\else \v{c}\fi{}i\ifmmode~\acute{c}\else
  \'{c}\fi{}}}]{malcolm-bitan}%
  \BibitemOpen
  \bibfield  {author} {\bibinfo {author} {\bibfnamefont {B.}~\bibnamefont
  {Roy}}, \bibinfo {author} {\bibfnamefont {M.~P.}\ \bibnamefont {Kennett}},
  \bibinfo {author} {\bibfnamefont {K.}~\bibnamefont {Yang}},\ and\ \bibinfo
  {author} {\bibfnamefont {V.}~\bibnamefont {Juri\ifmmode \check{c}\else
  \v{c}\fi{}i\ifmmode~\acute{c}\else \'{c}\fi{}}},\ }\bibfield  {title}
  {\bibinfo {title} {From birefringent electrons to a marginal or non-fermi
  liquid of relativistic spin-$1/2$ fermions: An emergent superuniversality},\
  }\href {https://doi.org/10.1103/PhysRevLett.121.157602} {\bibfield  {journal}
  {\bibinfo  {journal} {Phys. Rev. Lett.}\ }\textbf {\bibinfo {volume} {121}},\
  \bibinfo {pages} {157602} (\bibinfo {year} {2018})}\BibitemShut {NoStop}%
\bibitem [{\citenamefont {Nandkishore}\ and\ \citenamefont
  {Parameswaran}(2017)}]{rahul-sid}%
  \BibitemOpen
  \bibfield  {author} {\bibinfo {author} {\bibfnamefont {R.~M.}\ \bibnamefont
  {Nandkishore}}\ and\ \bibinfo {author} {\bibfnamefont {S.~A.}\ \bibnamefont
  {Parameswaran}},\ }\bibfield  {title} {\bibinfo {title} {Disorder-driven
  destruction of a non-fermi liquid semimetal studied by renormalization group
  analysis},\ }\href {https://doi.org/10.1103/PhysRevB.95.205106} {\bibfield
  {journal} {\bibinfo  {journal} {Phys. Rev. B}\ }\textbf {\bibinfo {volume}
  {95}},\ \bibinfo {pages} {205106} (\bibinfo {year} {2017})}\BibitemShut
  {NoStop}%
\bibitem [{\citenamefont {Mandal}\ and\ \citenamefont
  {Nandkishore}(2018)}]{ips-rahul}%
  \BibitemOpen
  \bibfield  {author} {\bibinfo {author} {\bibfnamefont {I.}~\bibnamefont
  {Mandal}}\ and\ \bibinfo {author} {\bibfnamefont {R.~M.}\ \bibnamefont
  {Nandkishore}},\ }\bibfield  {title} {\bibinfo {title} {Interplay of coulomb
  interactions and disorder in three-dimensional quadratic band crossings
  without time-reversal symmetry and with unequal masses for conduction and
  valence bands},\ }\href {https://doi.org/10.1103/PhysRevB.97.125121}
  {\bibfield  {journal} {\bibinfo  {journal} {Phys. Rev. B}\ }\textbf {\bibinfo
  {volume} {97}},\ \bibinfo {pages} {125121} (\bibinfo {year}
  {2018})}\BibitemShut {NoStop}%
\bibitem [{\citenamefont {Dalidovich}\ and\ \citenamefont {Lee}(2013)}]{denis}%
  \BibitemOpen
  \bibfield  {author} {\bibinfo {author} {\bibfnamefont {D.}~\bibnamefont
  {Dalidovich}}\ and\ \bibinfo {author} {\bibfnamefont {S.-S.}\ \bibnamefont
  {Lee}},\ }\bibfield  {title} {\bibinfo {title} {Perturbative non-fermi
  liquids from dimensional regularization},\ }\href
  {https://doi.org/10.1103/PhysRevB.88.245106} {\bibfield  {journal} {\bibinfo
  {journal} {Phys. Rev. B}\ }\textbf {\bibinfo {volume} {88}},\ \bibinfo
  {pages} {245106} (\bibinfo {year} {2013})}\BibitemShut {NoStop}%
\bibitem [{\citenamefont {Mandal}\ and\ \citenamefont
  {Lee}(2015)}]{ips-uv-ir1}%
  \BibitemOpen
  \bibfield  {author} {\bibinfo {author} {\bibfnamefont {I.}~\bibnamefont
  {Mandal}}\ and\ \bibinfo {author} {\bibfnamefont {S.-S.}\ \bibnamefont
  {Lee}},\ }\bibfield  {title} {\bibinfo {title} {Ultraviolet/infrared mixing
  in non-fermi liquids},\ }\href {https://doi.org/10.1103/PhysRevB.92.035141}
  {\bibfield  {journal} {\bibinfo  {journal} {Phys. Rev. B}\ }\textbf {\bibinfo
  {volume} {92}},\ \bibinfo {pages} {035141} (\bibinfo {year}
  {2015})}\BibitemShut {NoStop}%
\bibitem [{\citenamefont {Mandal}(2016)}]{ips-uv-ir2}%
  \BibitemOpen
  \bibfield  {author} {\bibinfo {author} {\bibfnamefont {I.}~\bibnamefont
  {Mandal}},\ }\bibfield  {title} {\bibinfo {title} {{UV/IR Mixing In Non-Fermi
  Liquids: Higher-Loop Corrections In Different Energy Ranges}},\ }\href
  {https://doi.org/10.1140/epjb/e2016-70509-4} {\bibfield  {journal} {\bibinfo
  {journal} {Eur. Phys. J. B}\ }\textbf {\bibinfo {volume} {89}},\ \bibinfo
  {pages} {278} (\bibinfo {year} {2016})}\BibitemShut {NoStop}%
\bibitem [{\citenamefont {Mandal}(2020)}]{ips-nfl-u1}%
  \BibitemOpen
  \bibfield  {author} {\bibinfo {author} {\bibfnamefont {I.}~\bibnamefont
  {Mandal}},\ }\bibfield  {title} {\bibinfo {title} {Critical fermi surfaces in
  generic dimensions arising from transverse gauge field interactions},\ }\href
  {https://doi.org/10.1103/PhysRevResearch.2.043277} {\bibfield  {journal}
  {\bibinfo  {journal} {Phys. Rev. Research}\ }\textbf {\bibinfo {volume}
  {2}},\ \bibinfo {pages} {043277} (\bibinfo {year} {2020})}\BibitemShut
  {NoStop}%
\bibitem [{Note1()}]{Note1}%
  \BibitemOpen
  \bibinfo {note} {There are two possible contractions (matching one $\protect
  \tilde \psi ^{\protect \dag }$ with one $\protect \tilde \psi $) of the
  interaction term in Eq.~\protect \textup {\hbox {\mathsurround \z@ \protect
  \normalfont (\ignorespaces \ref {action}\unskip \@@italiccorr )}} that gives
  the fermion self-energy. In the real space, the `Hartree' contractions are of
  the form $\langle \psi ^{\protect \dag } (t,x) \protect \,\psi (t,x) \rangle
  \protect \, \psi ^{\protect \dag } (t',{x'})\protect \, \psi (t',{x'})$, and
  correspond to tadpole diagrams. These simply shift the overall chemical
  potential, and can be ignored (since we assume that the renormalized chemical
  potential is at the band-crossing point). However, the `exchange'
  contractions contribute (cannot be ignored). These latter contributions are
  of the form $\langle \psi ^{\protect \dag } (t,x) \protect \,\psi (t',{x'})
  \rangle \protect \, \psi ^{\protect \dag }(t',{x'}) \protect \,\psi (t,x)$,
  and $\langle \psi ^{\protect \dag } (t',x') \protect \,\psi (t,{x}) \rangle
  \protect \, \psi ^{\protect \dag }(t,{x}) \protect \,\psi (t',x')$. Due to
  two ways of contractions, we get a factor of 2.}\BibitemShut {Stop}%
\bibitem [{\citenamefont {Pimenov}\ \emph {et~al.}(2018)\citenamefont
  {Pimenov}, \citenamefont {Mandal}, \citenamefont {Piazza},\ and\
  \citenamefont {Punk}}]{ips-fflo}%
  \BibitemOpen
  \bibfield  {author} {\bibinfo {author} {\bibfnamefont {D.}~\bibnamefont
  {Pimenov}}, \bibinfo {author} {\bibfnamefont {I.}~\bibnamefont {Mandal}},
  \bibinfo {author} {\bibfnamefont {F.}~\bibnamefont {Piazza}},\ and\ \bibinfo
  {author} {\bibfnamefont {M.}~\bibnamefont {Punk}},\ }\bibfield  {title}
  {\bibinfo {title} {Non-fermi liquid at the fflo quantum critical point},\
  }\href {https://doi.org/10.1103/PhysRevB.98.024510} {\bibfield  {journal}
  {\bibinfo  {journal} {Phys. Rev. B}\ }\textbf {\bibinfo {volume} {98}},\
  \bibinfo {pages} {024510} (\bibinfo {year} {2018})}\BibitemShut {NoStop}%
\bibitem [{\citenamefont {Eberlein}\ \emph {et~al.}(2016)\citenamefont
  {Eberlein}, \citenamefont {Mandal},\ and\ \citenamefont
  {Sachdev}}]{ips-subir}%
  \BibitemOpen
  \bibfield  {author} {\bibinfo {author} {\bibfnamefont {A.}~\bibnamefont
  {Eberlein}}, \bibinfo {author} {\bibfnamefont {I.}~\bibnamefont {Mandal}},\
  and\ \bibinfo {author} {\bibfnamefont {S.}~\bibnamefont {Sachdev}},\
  }\bibfield  {title} {\bibinfo {title} {Hyperscaling violation at the
  ising-nematic quantum critical point in two-dimensional metals},\ }\href
  {https://doi.org/10.1103/PhysRevB.94.045133} {\bibfield  {journal} {\bibinfo
  {journal} {Phys. Rev. B}\ }\textbf {\bibinfo {volume} {94}},\ \bibinfo
  {pages} {045133} (\bibinfo {year} {2016})}\BibitemShut {NoStop}%
\bibitem [{\citenamefont {Mandal}(2017)}]{ips-c2}%
  \BibitemOpen
  \bibfield  {author} {\bibinfo {author} {\bibfnamefont {I.}~\bibnamefont
  {Mandal}},\ }\bibfield  {title} {\bibinfo {title} {Scaling behaviour and
  superconducting instability in anisotropic non-fermi liquids},\ }\href
  {https://doi.org/10.1016/j.aop.2016.11.009} {\bibfield  {journal} {\bibinfo
  {journal} {Annals of Physics}\ }\textbf {\bibinfo {volume} {376}},\ \bibinfo
  {pages} {89–107} (\bibinfo {year} {2017})}\BibitemShut {NoStop}%
\bibitem [{\citenamefont {Mandal}\ and\ \citenamefont
  {Ziegler}(2021)}]{ips-klaus}%
  \BibitemOpen
  \bibfield  {author} {\bibinfo {author} {\bibfnamefont {I.}~\bibnamefont
  {Mandal}}\ and\ \bibinfo {author} {\bibfnamefont {K.}~\bibnamefont
  {Ziegler}},\ }\bibfield  {title} {\bibinfo {title} {Robust quantum transport
  at particle-hole symmetry},\ }\href
  {https://doi.org/10.1209/0295-5075/ac1a25} {\bibfield  {journal} {\bibinfo
  {journal} {EPL (Europhysics Letters)}\ }\textbf {\bibinfo {volume} {135}},\
  \bibinfo {pages} {17001} (\bibinfo {year} {2021})}\BibitemShut {NoStop}%
\bibitem [{\citenamefont {Mandal}\ and\ \citenamefont
  {Freire}(2021)}]{ips-freire1}%
  \BibitemOpen
  \bibfield  {author} {\bibinfo {author} {\bibfnamefont {I.}~\bibnamefont
  {Mandal}}\ and\ \bibinfo {author} {\bibfnamefont {H.}~\bibnamefont
  {Freire}},\ }\bibfield  {title} {\bibinfo {title} {Transport in the non-fermi
  liquid phase of isotropic luttinger semimetals},\ }\href
  {https://doi.org/10.1103/PhysRevB.103.195116} {\bibfield  {journal} {\bibinfo
   {journal} {Phys. Rev. B}\ }\textbf {\bibinfo {volume} {103}},\ \bibinfo
  {pages} {195116} (\bibinfo {year} {2021})}\BibitemShut {NoStop}%
\bibitem [{\citenamefont {Freire}\ and\ \citenamefont
  {Mandal}(2021)}]{ips-freire2}%
  \BibitemOpen
  \bibfield  {author} {\bibinfo {author} {\bibfnamefont {H.}~\bibnamefont
  {Freire}}\ and\ \bibinfo {author} {\bibfnamefont {I.}~\bibnamefont
  {Mandal}},\ }\bibfield  {title} {\bibinfo {title} {Thermoelectric and thermal
  properties of the weakly disordered non-fermi liquid phase of luttinger
  semimetals},\ }\href
  {https://doi.org/https://doi.org/10.1016/j.physleta.2021.127470} {\bibfield
  {journal} {\bibinfo  {journal} {Physics Letters A}\ }\textbf {\bibinfo
  {volume} {407}},\ \bibinfo {pages} {127470} (\bibinfo {year}
  {2021})}\BibitemShut {NoStop}%
\bibitem [{\citenamefont {Roy}\ and\ \citenamefont {Juri\ifmmode \check{c}\else
  \v{c}\fi{}i\ifmmode~\acute{c}\else \'{c}\fi{}}(2020)}]{bitan-vladimir}%
  \BibitemOpen
  \bibfield  {author} {\bibinfo {author} {\bibfnamefont {B.}~\bibnamefont
  {Roy}}\ and\ \bibinfo {author} {\bibfnamefont {V.}~\bibnamefont {Juri\ifmmode
  \check{c}\else \v{c}\fi{}i\ifmmode~\acute{c}\else \'{c}\fi{}}},\ }\bibfield
  {title} {\bibinfo {title} {Relativistic non-fermi liquid from interacting
  birefringent fermions: A robust superuniversality},\ }\href
  {https://doi.org/10.1103/PhysRevResearch.2.012047} {\bibfield  {journal}
  {\bibinfo  {journal} {Phys. Rev. Research}\ }\textbf {\bibinfo {volume}
  {2}},\ \bibinfo {pages} {012047} (\bibinfo {year} {2020})}\BibitemShut
  {NoStop}%
\bibitem [{\citenamefont {Mandal}(2018)}]{ips-qbt-sc}%
  \BibitemOpen
  \bibfield  {author} {\bibinfo {author} {\bibfnamefont {I.}~\bibnamefont
  {Mandal}},\ }\bibfield  {title} {\bibinfo {title} {Fate of superconductivity
  in three-dimensional disordered luttinger semimetals},\ }\href
  {https://doi.org/10.1016/j.aop.2018.03.004} {\bibfield  {journal} {\bibinfo
  {journal} {Annals of Physics}\ }\textbf {\bibinfo {volume} {392}},\ \bibinfo
  {pages} {179–195} (\bibinfo {year} {2018})}\BibitemShut {NoStop}%
\end{thebibliography}%
\end{document}